\documentclass[aps,onecolumn,preprintnumbers,nofootinbib,superscriptaddress]{revtex4}
\usepackage{eurosym}
\usepackage{amsmath}
\usepackage{graphicx}
\usepackage{amsfonts}
\usepackage{array}
\usepackage{amsthm}
\usepackage{bm}
\usepackage{palatino}
\usepackage{mathpazo}
\usepackage{supertabular}
\usepackage{subfig}
\usepackage[breaklinks]{hyperref}
\usepackage[font=small,labelfont=bf,
justification=justified,
format=plain]{caption}
\usepackage[font=small,labelfont=bf,justification=justified]{caption}
\usepackage{color}
\usepackage{booktabs}
\usepackage{multirow}
\usepackage[labelfont=bf,labelsep=quad,justification=justified]{caption}
\newcommand{\be}{\begin{equation}}
\newcommand{\ee}{\end{equation}}
\newcommand{\ba}{\begin{eqnarray}}
\newcommand{\ea}{\end{eqnarray}}
\newcommand{\bal}{\begin{align}}
\newcommand{\eal}{\end{align}}

\newcommand{\bw}{\begin{widetext}}
\newcommand{\ew}{\end{widetext}}

\begin{document}

\title{Distinguishing a Kerr-like black hole and a naked singularity in perfect fluid dark
matter via precession frequencies}
\author{Muhammad Rizwan}
\email{m.rizwan@sns.nust.edu.pk}
\affiliation{Department of Mathematics, School of Natural Sciences (SNS), National
University of Sciences and Technology (NUST), H-12, Islamabad, Pakistan}
\affiliation{Faculty of Engineering and Computer Sciences, National University of Modern
Languages, H-9, Islamabad, Pakistan}
\author{Mubasher Jamil}
%\email{jamil.camp@gmail.com}
\affiliation{Department of Mathematics, School of Natural Sciences (SNS), National
University of Sciences and Technology (NUST), H-12, Islamabad, Pakistan}
\author{Kimet Jusufi}
\email{kimet.jusufi@unite.edu.mk}
\affiliation{Physics Department, State University of Tetovo, Ilinden Street nn, 1200,
Tetovo, Macedonia}
\affiliation{Institute of Physics, Faculty of Natural Sciences and Mathematics, Ss. Cyril
and Methodius University, Arhimedova 3, 1000 Skopje, Macedonia}

\begin{abstract}
We study a Kerr-like black hole and naked singularity in perfect fluid dark
matter (PFDM). The critical value of spin parameter $a_c$ is presented to
differentiate the black hole from naked singularity. It is seen that for any
fixed value of dark matter parameter $\alpha$ the rotating object is black
hole if $a\leq a_c$ and naked singularity if $a>a_c$. Also for $%
-2\leq\alpha<2/3$ the size of the black hole horizons decrease whereas for $%
2/3<\alpha$ it increases. We also study spin precession frequency of a test
gyroscope attached to stationary observer to differentiate a black hole from
naked singularity in PFDM. For the black hole, spin
precession frequency blows up as the observer reaches the central object while
for naked singularity it remains finite except at the ring singularity. Moreover,
we study Lense-Thirring precession for a Kerr-like black hole and geodetic
precession for Schwarzschild black hole in PFDM. To this end, we have calculated the Kepler frequency (KF), the vertical epicyclic frequency (VEF), and the nodal plane precession frequency (NPPF). Our results show that, the PFDM parameter $\alpha$ significantly affects those frequencies. This difference can be used by astrophysical observations in the near future to shed
some light on the nature of dark matter. 
\end{abstract}

\maketitle


%\newpage

\section{Introduction}

It is widely believed that the center of nearly every galaxy contains a
supermassive black hole. In particular, the increase of the astronomical
observations in recent years strongly indicates the presence of a
supermassive black hole at the center of our galaxy (Sgr A*). According to
the current model of cosmology, dark matter makes up about $27\%$ of the matter-energy composition of the
Universe, although, as of today, there is no direct experimental detection
of dark matter. Nevertheless, indirect experimental observations strongly
suggest that dark matter reveal its presence in many astrophysical
phenomena, especially important, in this context, are the problem of
galactic rotation curves \cite{dm1}, the galaxy clusters dynamics \cite{dm2}%
, while further evidence for dark matter comes from measurements on
cosmological scales of anisotropies in the cosmic microwave background
through PLANCK \cite{dm3}.

Therefore, it is extremely important to study the black hole physics in the
presence of dark matter. Li and Yang investigated the possibility of the static black hole immersed in dark matter \cite{SCindark}. Their model of dark matter is based on a single parameter $\alpha$ which is the limitation of the model. Furthermore their model corresponds to a specific case studied for the first time by Kiselev \cite{kiselev}. In particular the logarithmic dependence was introduced to explain the asymptotic rotation curves for the dark matter in terms of the of quintessential matter at large distances, i.e. in the halo dominated region. That being said, one possible limitation of this model is the fact that no interaction between the dark matter and other fields (say, dark energy field) is assumed. One can certainly modify the distribution of dark matter in a galaxy by considering an interaction between those fields. In other words, one may consider a more general scenario with a surrounding matter given as a combination of more complicated fields with more dark matter parameters.

Quite recently, a new Kerr black hole solution with
the dark matter effects has been reported in the literature \cite{KerrDM}.
This solution modifies the Kerr metric due to the presence of dark matter
encoded via the PFDM (PFDM) $\alpha$ which, among
other things, implies a modification of the ergosphere structure of the
black hole. This solution allows to study the effect of PFDM in different astrophysical problems. Very recently, this solution was used in \cite{shadow1,shadow2}, to
study the effect of PFDM and cosmological constant on the
size of black hole shadow, deflection angle, as well as the black hole emission rate which is related to the idea that for a far distant observer located at infinity the observed area of the black hole shadow approximately equals to the high energy absorption cross section. 
According to the general theory of relativity, there is a rotational
dragging of inertial frames near the presence of a rotating black hole spacetimes known
as a Lense-Thirring precession (LT) \cite{lt1,lt2,lt3}. Basically, we can
explore the dragging effects with the help of a gyroscope (or a test gyro)
using the fact that a gyroscope tends to keep its spin axis rigidly pointed
in a fixed direction in space, say fixed relative to a given star. In a
rotating spacetime, due to the frame dragging effects, it is shown that the
precession of the gyroscope frequency is proportional to the spin parameter
of the rotating object and inversely proportional to cube of the distance
from the central object. In addition to that, there is a second effect
related to the gyroscopic precession due to the spacetime curvature which is
known as a geodetic precession \cite{gdp}. LT precession of a test gyroscope
has been extensively studied in recent years, along this line of thought, in 
\cite{rotw} authors study the LT precession frequency in a rotating
traversable Teo wormhole, in \cite{strong} the frame-dragging effect in a
strong gravity regime is considered, and references therein. It is worth
noting that, great effort has been made to actually test the frame dragging
effect and geodetic effect in the Earth's gravitational field by the Gravity
Probe B experiment \cite{Everitt}. 

The concept of the spacetime singularity is well known in general relativity
mainly due to the famous Penrose-Hawking singularity theorem. %A fundamental difference between the naked singularity
%and regular spacetimes is related to the fact that a naked singularity is a gravitational singularity without an event
%horizon. 
A naked singularity on the other hand is defined as a gravitational singularity without an event horizon. According to the cosmic censorship conjecture, spacetime
singularities that arise in gravitational collapse are always hidden inside
of black holes and therefore can not be observed in nature \cite%
{nsing1,nsing2}. Whether naked singularities exist or not is an open
question, however, one can naturally raise the following intriguing question
concerning the nature of the final product of gravitational collapse: How
can we distinguish a naked singularity from a black hole? In this context,
the problem of naked singularities has attracted a great interest in recent
years \cite{new1,new2,new3}.

From the no-hair theorem we know that a Kerr solution is completely
characterized by the black hole mass $M$ and the black hole angular momentum 
$J$. If the following condition $M \geq a$ holds, where the angular momentum
parameter $a$ is defined by the angular momentum per unit mass, then the
Kerr solution represents a black hole. On the other hand, if $M<a$, then a
naked singularity is recovered. In a very interesting work, Chakraborty et
al \cite{NS,LTkerr} argued that one can basically use the spin precession
frequency of a test gyroscope attached to both static and stationary
observers, to distinguish black holes from naked singularities. Afterwards,
a new spin forward to this idea was put by Rizwan et al \cite{RMA} who
studied the problem of distinguishing a rotating Kiselev black hole from a
naked singularity using spin precession of test gyroscope. In some recent
papers \cite{bambi0}, authors study the idea of distinguishing black holes
and naked singularities with iron line spectroscopy, rotating naked
singularities are studied in the context of gravitational lensing \cite%
{galin}, while in \cite{kimet}, authors study the problem of distinguishing
rotating naked singularities from Kerr-like wormholes by their deflection
angles of massive particles.

It is known that the process of matter accretion towards rotating neutron
stars and black holes is followed by the emission of electromagnetic waves,
mainly X and gamma-rays \cite{xray1}. 
%Experimental observation of the power
%spectrum of the time series of the X-rays suggest the quasi-periodic
%oscillations (QPOs) phenomena \cite{xray2}.
The quasi-periodic oscillations phenomena (QPOs) is linked with high frequency X-ray binaries \cite{stella1,stella2}. In particular, there are known the high-frequency (HF) quasi-periodic oscillations (QPOs) and three types	of low-frequency (LF) QPOs. It is quite amazing that the LT effect can be
linked with this phenomena and perhaps to explain the QPOs of accretion
disks around rotating black holes, provided the disk is slightly misaligned
with the equatorial plane of the BH \cite{xray2}. 

%However, for a more realistic model one has to take into account the dark
%matter effect. 
In the present paper, firstly we shall
examine the critical value of spin parameter $a_c$ to differentiate the
Kerr-like black hole from a naked singularity with PFDM. Then, we shall
calculate the spin precession frequency of a test gyroscope attached to
stationary observer to differentiate a Kerr-like black hole from a naked
singularity with PFDM.

The outline of this paper is as follows. In Section II we determine the
critical value of spin parameter to differentiate a Kerr-like black hole
from from naked singularities in PFDM. In Section III, we calculate the spin
precession frequency of a test gyroscope in Kerr-like black hole with PFDM,
in particular we examine in detail the LT-precession of a gyroscope in
Kerr-like black hole with PFDM. In Section IV, we specialize our results to
elaborate the geodetic precession in Schwarzschild black hole spacetime in
PFDM. In Section V, we shall focus on the problem of distinguishing black
holes from naked singularities. In Section VI, we study the effect of PFDM on the KF, the VEF, and the NPPF.  Section VII is devoted for some concluding remarks.
\newpage

\section{Kerr-like black hole in perfect fluid dark matter\label{secGE}}

The line element of the Kerr-like black hole in PFDM is
given as \cite{KerrDM}

\begin{eqnarray}
ds^{2} &=&-\left( 1-\frac{2Mr-\alpha r\ln \left( \frac{r}{|\alpha |}\right) 
}{\Sigma }\right) dt^{2}+\frac{\Sigma }{\Delta }dr^{2}+\Sigma d\theta ^{2}-2a%
\frac{\left( 2Mr-\alpha r\ln \left( \frac{r}{|\alpha |}\right) \right) }{%
\Sigma }dtd\phi  \notag  \label{LE} \\
&&+\sin ^{2}\theta \left( r^{2}+a^{2}+a^{2}\sin ^{2}\theta \frac{2Mr-\alpha
r\ln \left( \frac{r}{|\alpha |}\right) }{\Sigma }\right) ,
\end{eqnarray}%
where%
\begin{equation}
\Delta \equiv r^{2}-2Mr+a^{2}+\alpha r\ln \left( \frac{r}{|\alpha |}\right) ,\text{
}\Sigma\equiv r^{2}+a^{2}\cos ^{2}\theta .
\end{equation}%
Here $M$ and $a$ are mass and angular momentum per unit mass, parameters of the black hole. Using
Komar integral, total mass of black hole $M_{T}$ interior to the surface $%
r=r_{0}$, and the corresponding angular momentum $J_{T}$ around the axis of
rotation of a stationary spacetime is obtained as 
\begin{equation}
M_{T}=M-\frac{\alpha \ln \left( \frac{r_{0}}{|\alpha |}\right) }{2ar_{0}}%
\left[ ar_{0}+\left( r_{0}^{2}+a^{2}\right) \tan ^{-1}\left( \frac{a}{r_{0}}%
\right) \right],
\end{equation}%
\begin{equation}
J_{T}=aM+\frac{\alpha }{4a^{2}r_{0}}\left[ \left( r_{0}^{2}+a^{2}\right)
^{2}\tan ^{-1}\left( \frac{a}{r_{0}}\right) -ar_{0}\left(
r_{0}^{2}+a^{2}\right) -2a^{3}r_{0}\ln \left( \frac{r_{0}}{|\alpha |}\right) %
\right].
\end{equation}

In the absence of PFDM ($\alpha =0$), the line element %
\eqref{LE} represents a Kerr black hole. The PDFM stress-energy tensor in
the standard orthogonal basis of the Kerr-like black hole can be written in
diagonal form $diag[-\rho,p_{r},p_{\theta},p_{\phi}]$ \cite{KerrDM}, where 
\begin{equation}
-\rho=p_{r}=\frac{\alpha r}{8\pi \Sigma^2}, p_{\theta}=p_{\phi}=\frac{\alpha
r}{8\pi \Sigma^2}\left(r-\frac{\Sigma}{2r}\right).
\end{equation}
The location of the black hole horizons can be obtained by solving the
horizon equation 
\begin{equation}  \label{Horizneq}
\Delta =r^{2}-2Mr+a^{2}+\alpha r\ln \left( \frac{r}{|\alpha |}\right) =0.
\end{equation}%
Note that depending on the choice of parameters $a$ and $\alpha$, %
\eqref{Horizneq} has no solution, one solution or two solutions. In each
case the line element \eqref{LE} represents naked singularity, extremal
black hole or black hole with inner $(r_-)$ and outer horizon $(r_+)$,
respectively. To find out the critical value (the maximum value of the
parameter for which \eqref{LE} can represent a black hole) of the spin
parameter $a_{c}$ in this section we express the black hole parameters and
the radial distance in the unit of black hole mass, that is, $a/M\rightarrow
a$, $\alpha /M\rightarrow \alpha $ and $r/M\rightarrow r$. Assuming the spin
parameter $a$ as a function of $r$ and $\alpha $ we can write 
\begin{equation}
a^{2}(r,\alpha )=2r-r^{2}-\alpha r\ln \left( \frac{r}{|\alpha |}\right) .
\label{a2}
\end{equation}%
Now to find the extreme value of the spin parameter $a$ we use the condition
of extrema of $a^{2}$, that is, $da^{2}/dr=0$, which yields 
\begin{equation}
f(r,\alpha )\equiv 2-2r-\alpha \ln \left( {\frac{r}{|\alpha |}}\right) {\
-\alpha }=0.  \label{extreme}
\end{equation}%
and for any fixed $\alpha $, 
\begin{equation}
\frac{df}{dr}=-2-\frac{\alpha }{r},
\end{equation}%
Note that, 
\begin{eqnarray}
\text{For any }\alpha &<&0,\text{ }\frac{df}{dr}>0\text{ for }0<r<-\frac{
\alpha }{2}\text{ and }\frac{df}{dr}<0\text{ for }-\frac{\alpha }{2}<r. \\
\text{For any }\alpha &>&0,\text{ }\frac{df}{dr}<0\text{ for all }r.
\end{eqnarray}
The above conditions show that the function $f(r,\alpha)$ behaves
differently for negative and positive values of $\alpha$. So we will discuss
these two cases separately.

\begin{figure}[!ht]
\centering
%\captionsetup{justification=centering}
 \minipage{0.50\textwidth} %
\includegraphics[width=8.2cm,height=5.4cm]{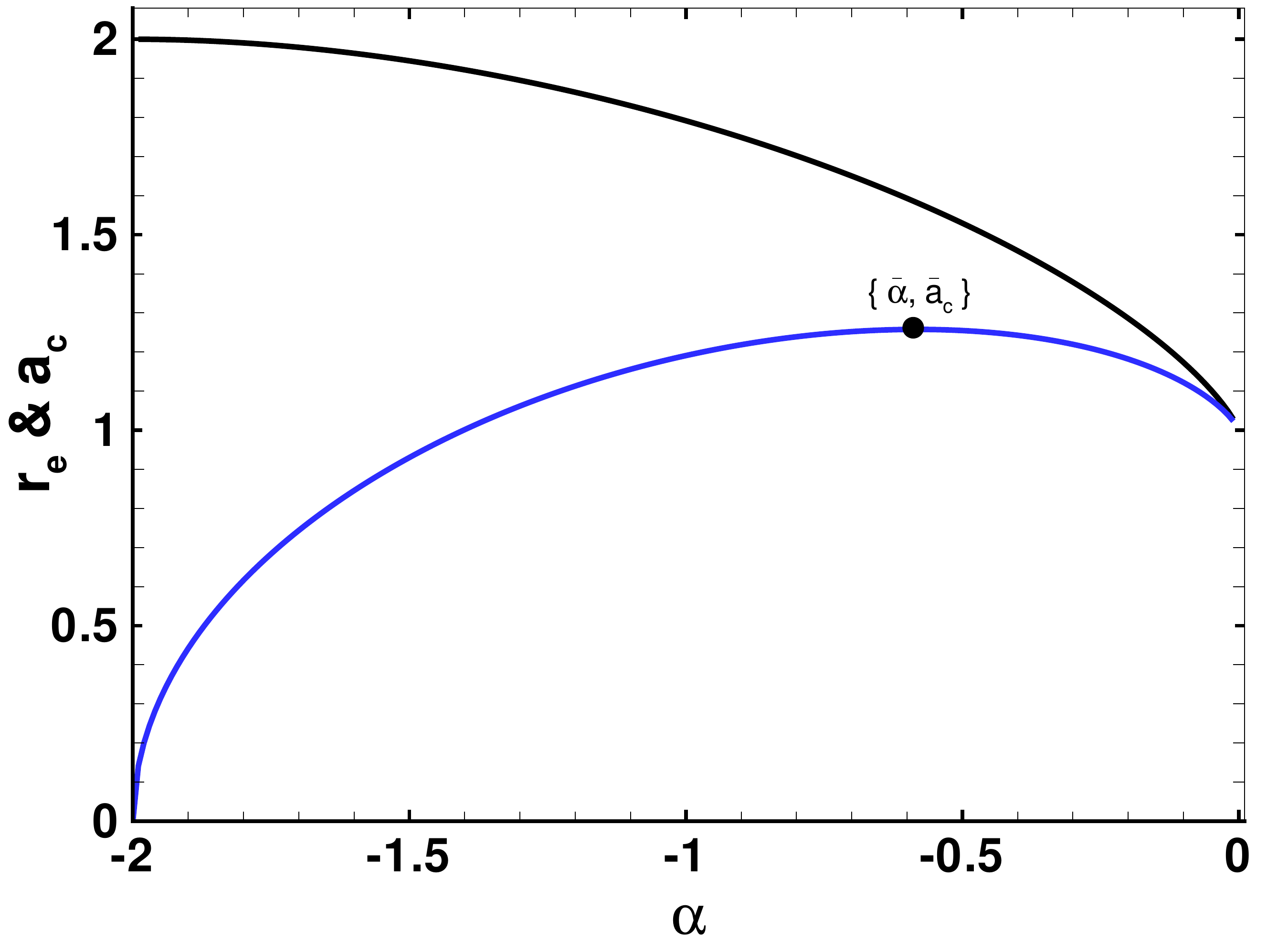}\newline
(a) \label{rea} \endminipage\hfill \minipage{0.50\textwidth} %
\includegraphics[width=8.0cm,height=5.6cm]{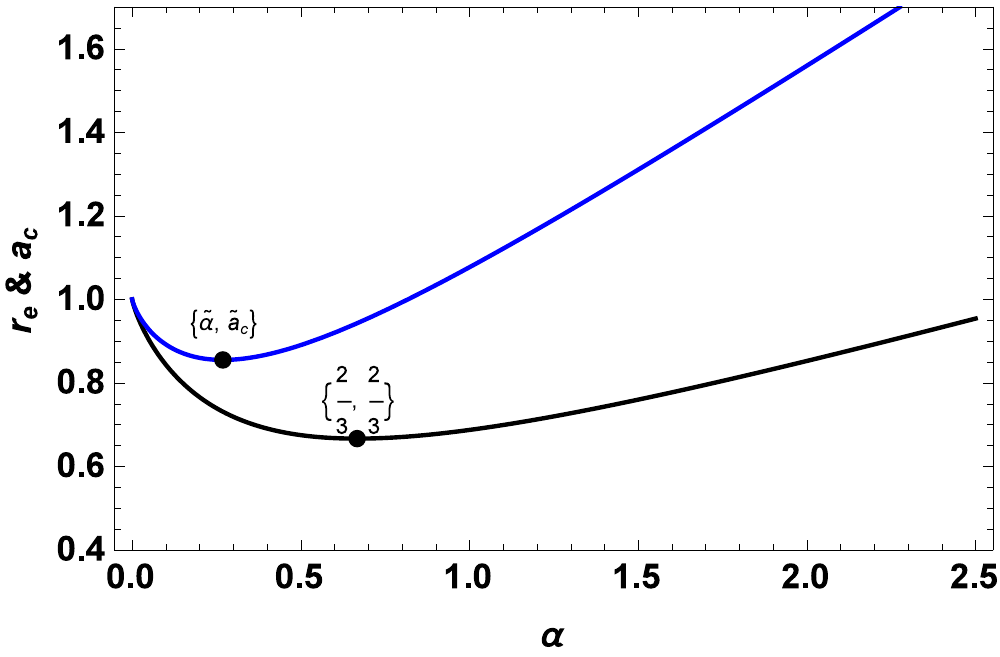}\newline
(b) \label{reb} \endminipage\hfill
\caption{{\protect\footnotesize The horizon of the extremal black hole $r_e$
(black line) and critical values of the spin parameter $a_c$ (blue line)
verse negative and positive $\protect\alpha$ are plotted in panel (a) and
(b). Here $\overline{\protect\alpha} \approx-0.581977$, $\overline{a}
_c\approx1.25776655499709$, $\tilde{\protect\alpha}=1/(1+e)$ and $\tilde{a}
_c\approx0.855$.}}
\label{reandac}
\end{figure}

\subsection{Negative $\alpha $}%\newline
For any fixed chosen $\alpha<0$, the function $f(r,\alpha)$ has
maxima at $r=-\alpha/{2}$. The function increases in the interval $0< r <-\alpha/2$
while decreases for $-\alpha/{2}<r$. Thus, depending on the value of $%
\alpha $, $f(r,\alpha)$ have no zero or have two zeros say $r_1$ and $r_2$
such that $r_1\leq -\frac{\alpha}{2} \leq r_2$. If $\alpha_\text{min}$ is
the minimum value for which $f(r_,\alpha)$ has a zero, than 
\begin{equation}
r_1=r_2 \quad \text{for} \quad \alpha=\alpha_\text{min}.
\end{equation}
That is, $r_1$ is a zero of $f(r,\alpha)$ of multiplicity $2$. Solving %
\eqref{extreme} for negative $\alpha$, we find that one zero of $f(r,\alpha)$
is 
\begin{equation}  \label{r1-}
r_1=\frac{\alpha }{2}{ProductLog}\left( {-2e^{-1+\frac{2}{\alpha }}}\right),
\end{equation}
where $ProductLog(x)$ is a Lambert \textit{W-}function. Now, if $r_1$ is
zero of $f(r,\alpha)$ of multiplicity $2$, then it must also be zero of $%
df/dr$, which gives 
\begin{equation}
\alpha_\text{min}=-\frac{2}{ln(2)}\approx-2.88539.
\end{equation}
Note that for any value of $\alpha$ in the range $\alpha_\text{min}\leq
\alpha<0$ the corresponding extremal value of $a$ are 
\begin{equation}
a^2=r_1(r_1+\alpha) \quad and \quad a^2=r_2(r_2+\alpha).
\end{equation}
As $r_1\leq -\alpha/2$, so in this case $a^2$ is negative which implies $r_1$
cannot be horizon of the extremal black hole and thus $r_2$ can be the
horizon of the extremal black hole. Further, for $\alpha=-2$, $r_2=2$ and $%
a=0$. If $\alpha$ is in the range $-\alpha_{min}\leq \alpha<2 $, the
cosponsoring solution $r_2$ gives $a^2$ negative. Thus, for negative $\alpha$
the line element \eqref{LE} represents a black hole only if $-2\leq\alpha<0$%
. Solving \eqref{extreme} numerically for $-2\leq \alpha<0 $ gives horizon
of the extremal black hole $r_2$ and henceforth will be represented by $r_e$.
Using $r_e$ in \eqref{extreme}, we get the critical value of spin parameter as follows 
\begin{equation}
a_c=\sqrt{r_e(r_e+\alpha)}.
\end{equation}
The graph of $r_e$ and $a_c$ for $-2\leq \alpha <0$ is plotted in FIG:\ref%
{reandac} (a) which shows that $r_e$ decreases with increasing $\alpha$,
while $a_c$ has its maximum value $\overline{a}_c\approx1.2577$ at 
$\overline{\alpha} \approx-0.58197$. If $\alpha$ is in the range, $%
-2\leq\alpha< \overline{\alpha}$ the critical value of spin parameter $a_c$
increases and if $\overline{\alpha}<\alpha<0$, then $a_c$ decreases.
Further, as $r_-\leq r_e\leq r_+$ so we can say that for $-2\leq \alpha <0$,
with increasing $\alpha$ the size of inner horizon $r_-$ decreases. 
\begin{figure}[th]
\centering
%\captionsetup{justification=centering}
 \minipage{0.31\textwidth} %
\includegraphics[width=2.3in,height=1.6in]{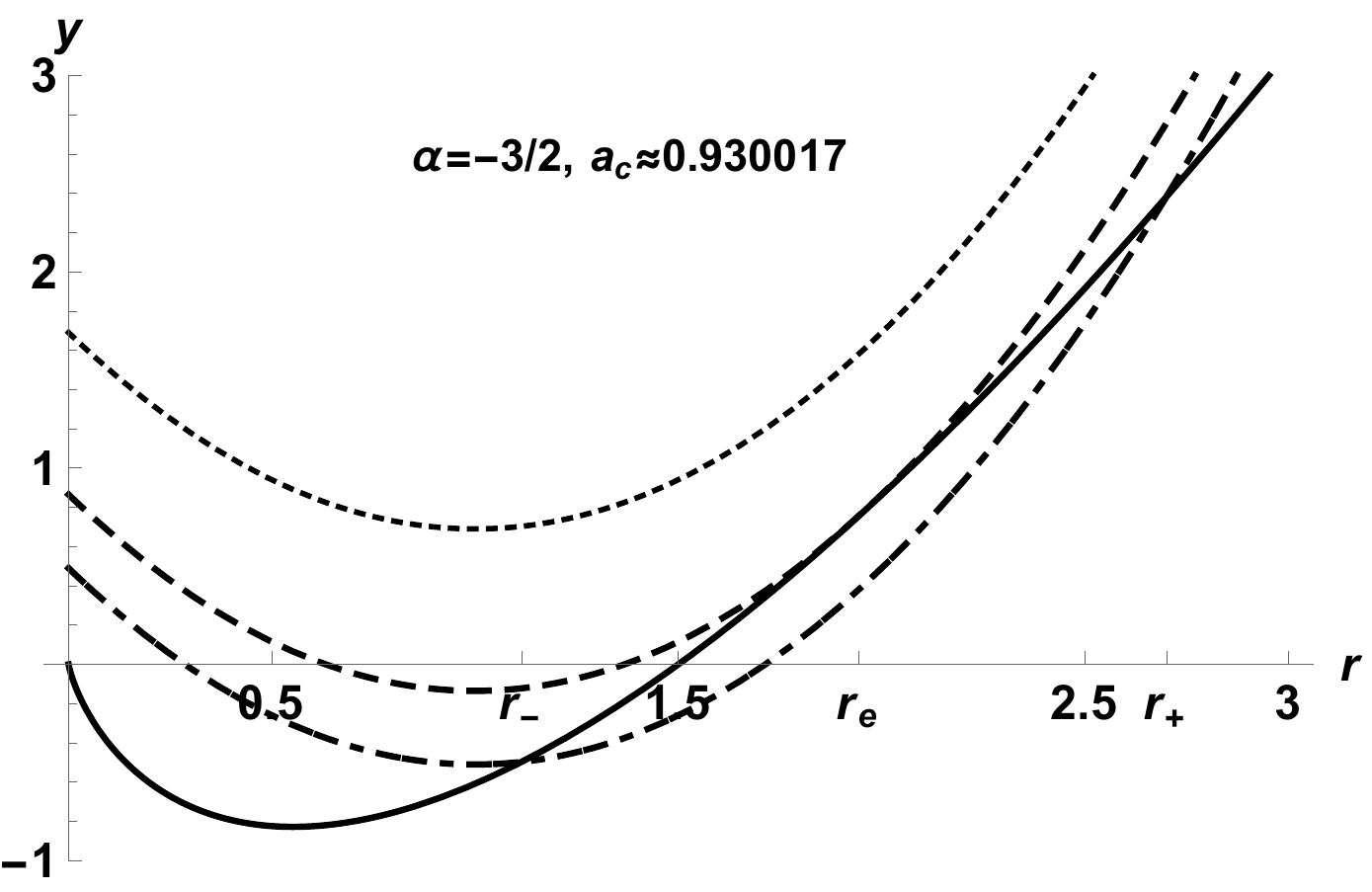}\newline
(a)  \label{a} \endminipage\hfill %
\minipage{0.31\textwidth} %
\includegraphics[width=2.3in,height=1.6in]{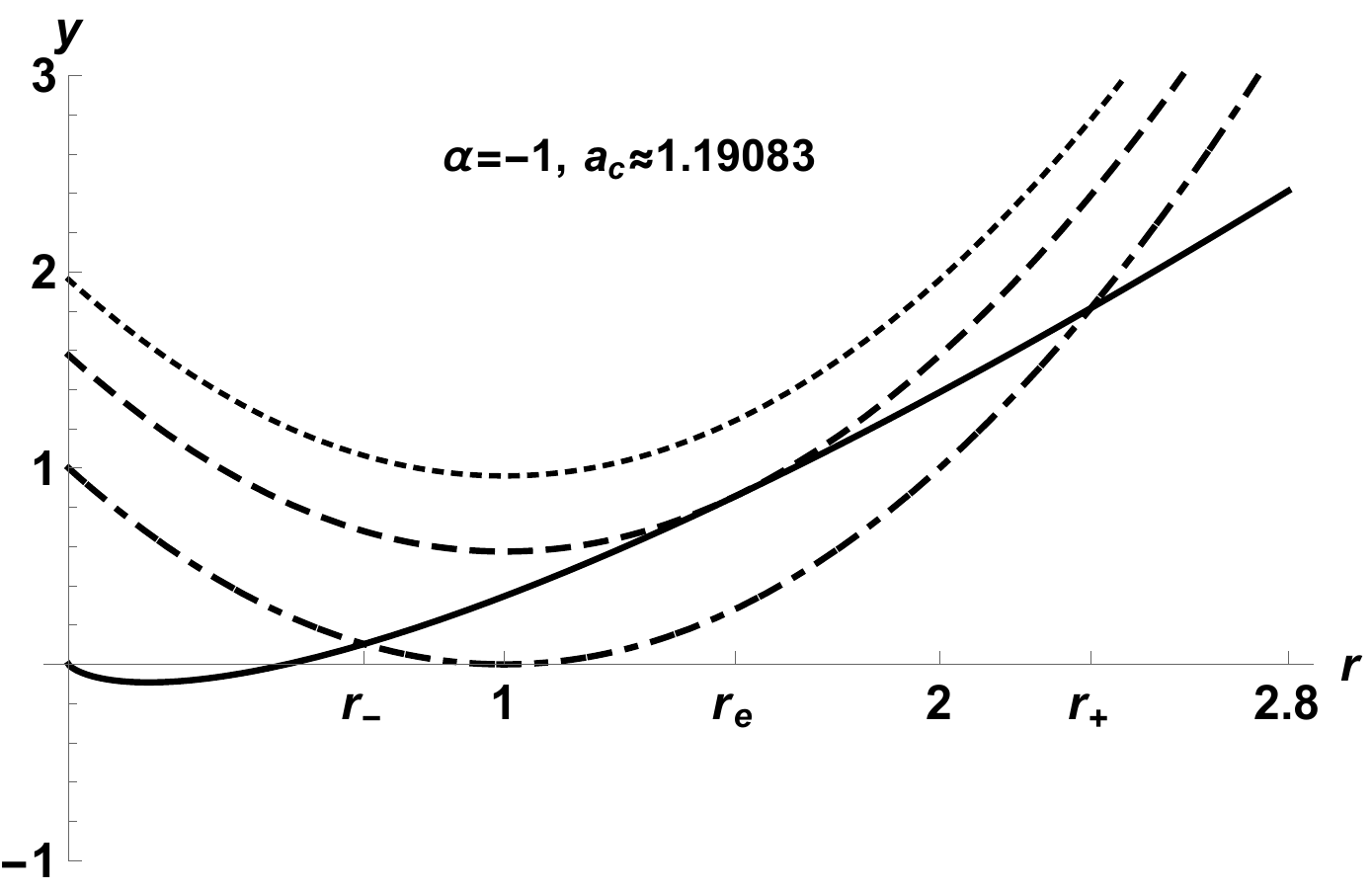}\newline
(b)  \label{b} \endminipage\hfill %
\minipage{0.31\textwidth} %
\includegraphics[width=2.3in,height=1.6in]{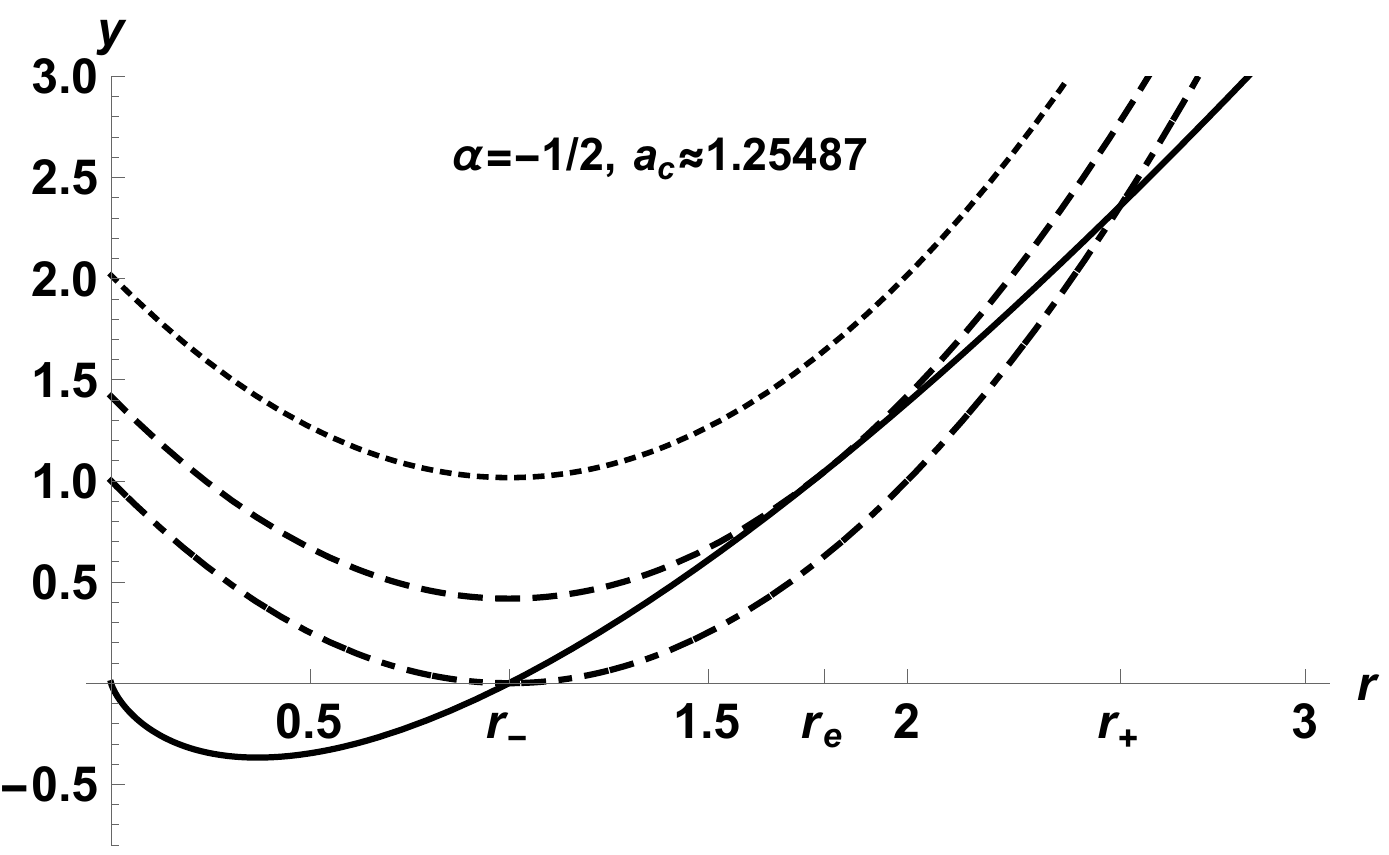}\newline
(c)  \label{c} \endminipage\hfill
\par
\minipage{0.31\textwidth} %
\includegraphics[width=2.3in,height=1.6in]{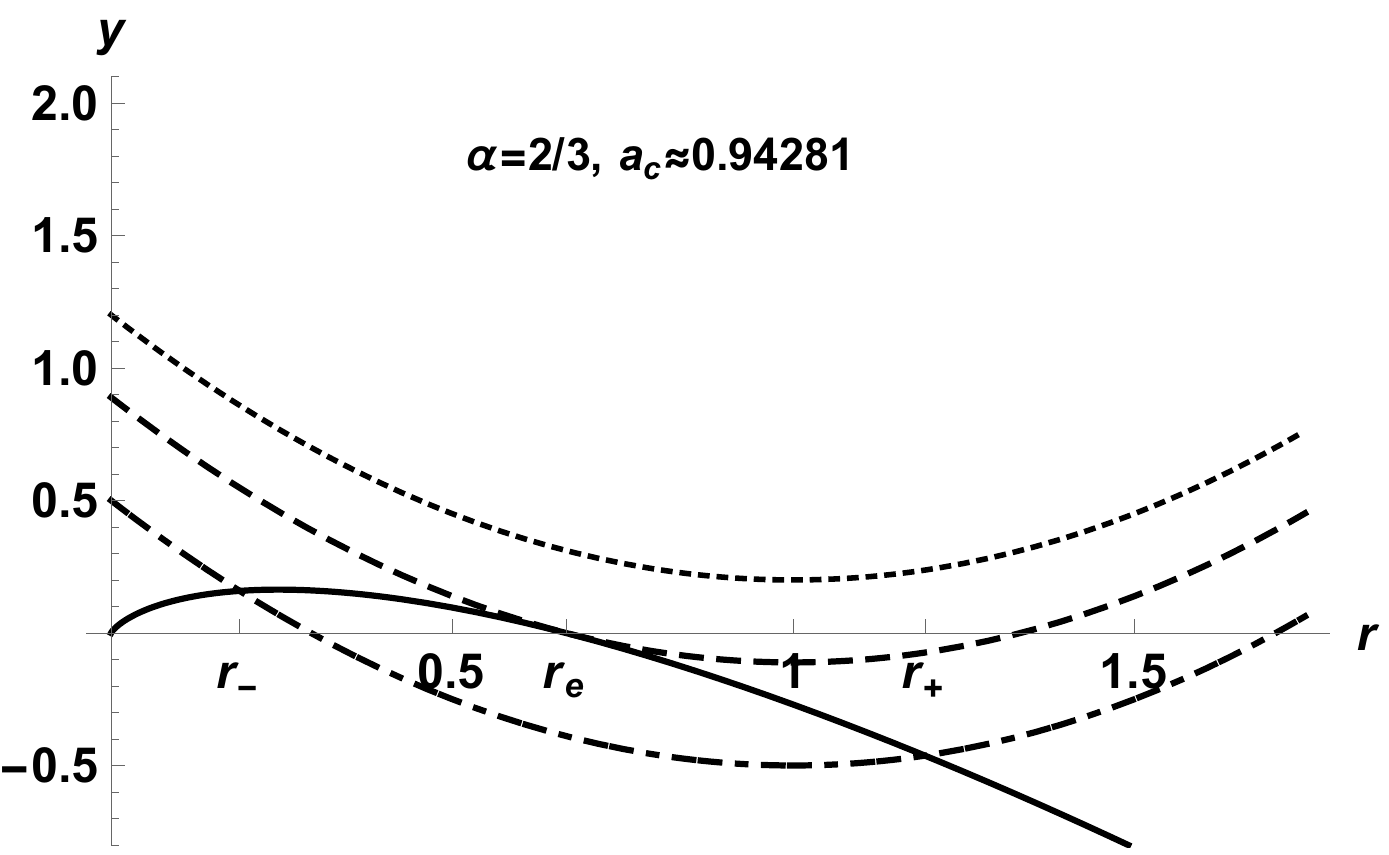}\newline
(d)  \label{6} \endminipage\hfill %
\minipage{0.31\textwidth} %
\includegraphics[width=2.3in,height=1.6in]{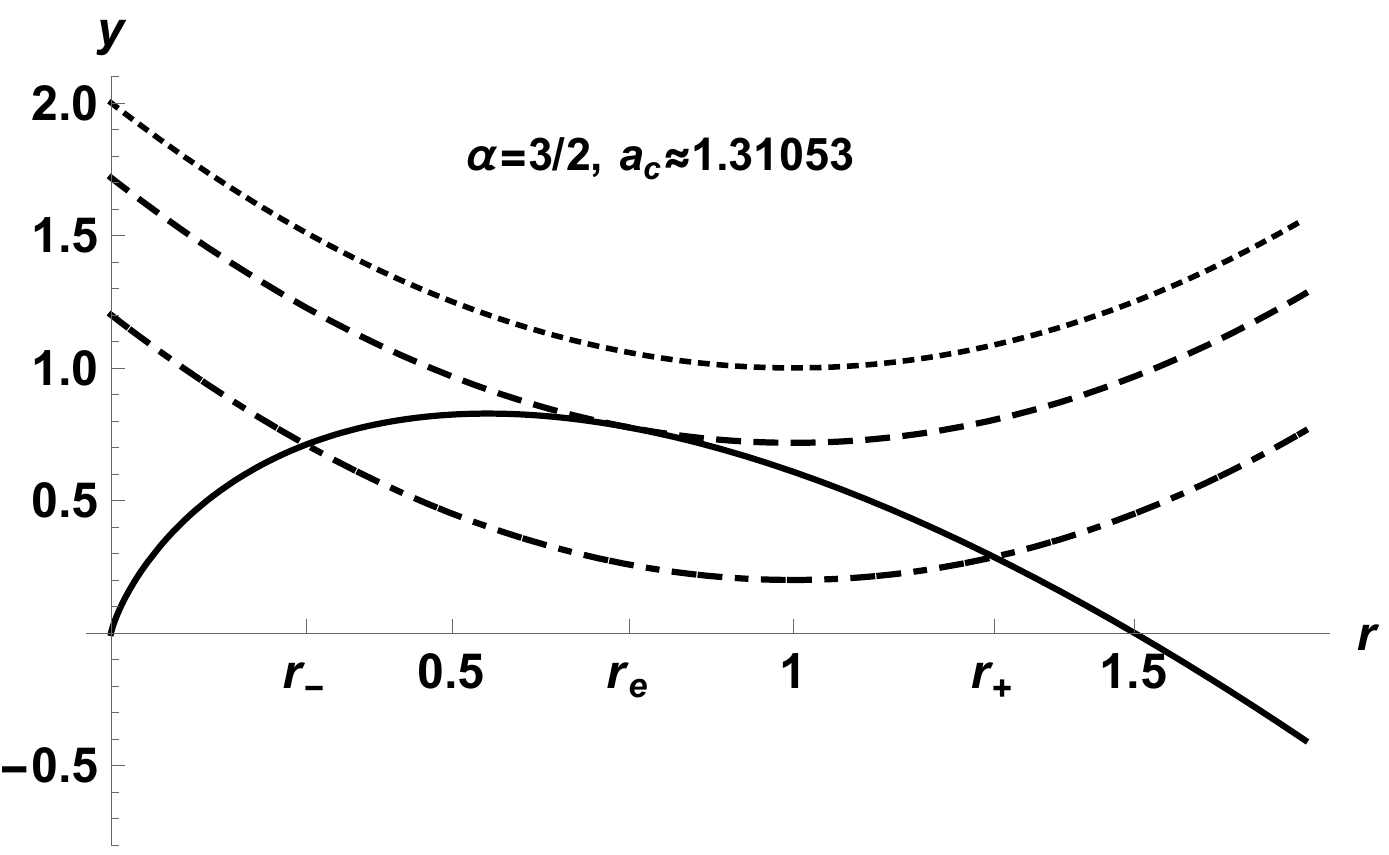}\newline
(e)  \label{e} \endminipage\hfill %
\minipage{0.31\textwidth} %
\includegraphics[width=2.3in,height=1.6in]{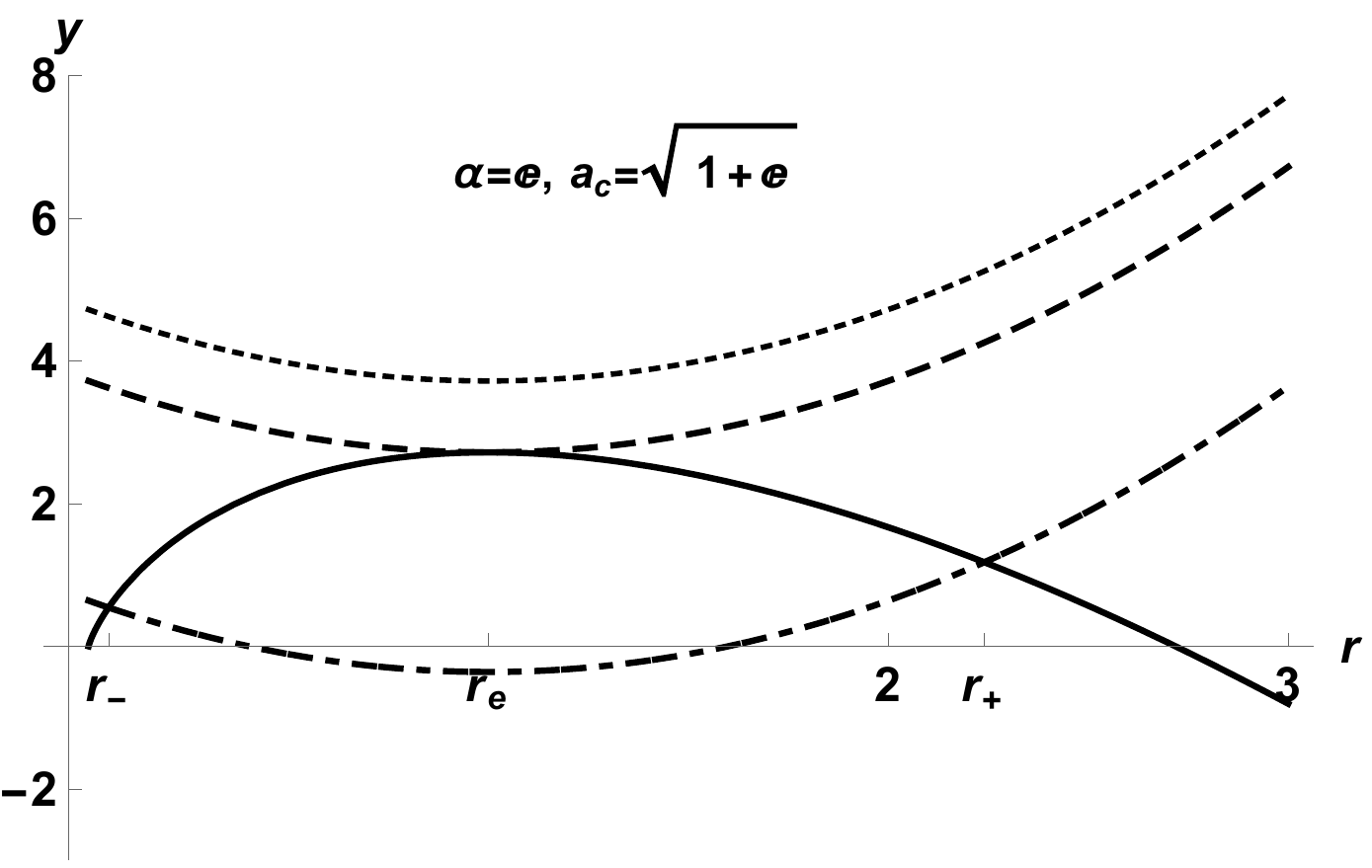}\newline
(f)  \label{f} \endminipage\hfill
\caption{{\protect\footnotesize The graphs of $y=r^{2}-2r+a^{2}$ for
different value of $a$, $a<a_{c}$ (dash-dotted line), $a=a_{c}$ (dashed
line) and $a>a_{c}$ (dotted line) and $y=-\protect\alpha rln(\frac{r}{| 
\protect\alpha |})$ (solid line) are plotted. For any $\protect\alpha $ and $%
a<a_{c}$, the point of intersection of \emph{solid lines} and \emph{\
dash-dotted parabolas} in (a)-(f), give the locations of inner horizons $%
r_{-}$ and event horizons $r_{+}$. For $a=a_{c}$, the point of intersection
of \emph{solid lines} and \emph{dashed parabolas} give the locations of
horizon of extremal black hole $r_{e}$. For $a>a_{c}$, the \emph{solid line}
and \emph{dotted line} does not intersect each other indicating that there
does not exist a black hole and the line element \eqref{LE} represents a
naked singularity.}}
\label{Fig4}
\end{figure}
\newline

\subsection{Case II: Positive $\alpha $}
To discuss the critical value $a_{c}$ for any positive $\alpha $, we first
find zeros of the function $f(r,\alpha )$. As for any chosen positive $%
\alpha $ the function $f(r,\alpha )$ is decreasing function of $r$ so it has
at most one zero. Further 
\begin{equation}
f(\overline{r}_{1},\alpha )=2>0\text{ \ with \ }\overline{r}_{1}=\frac{
\alpha }{2}ProductLog\left( \frac{2}{e}\right) ,
\end{equation}%
and 
\begin{equation}
f\left( \overline{r}_{2},\alpha \right) =-2\alpha e^{-1+\frac{2}{\alpha }}<0 
\text{ \ with \ }\overline{r}_{2}=\alpha e^{-1+\frac{2}{\alpha }},
\end{equation}%
By Intermediate value theorem we can conclude that for any $\alpha >0,$ $%
f\left( r,\alpha \right) $ has one zero (say $r_{e}$) such that $\overline{r}%
_{1}<r_{e}$ $<\overline{r}_{2}$. Solving \eqref{extreme} for $r$ yields 
\begin{equation}
r_{e}=\frac{\alpha }{2}{ProductLog}\left( {2e^{-1+\frac{2}{\alpha }}}\right)
.
\end{equation}%
and the corresponding extreme value of spin parameter $a_{c}$ is obtained as 
\begin{equation}
a=\sqrt{r_{e}\left( 2-r_{e}-\alpha \ln \left( \frac{r_{e}}{\alpha }\right)
\right) },
\end{equation}%
or 
\begin{equation}\label{ac}
a_{c}=\frac{\alpha }{2}\sqrt{ProductLog\left( 2e^{-1+\frac{2}{\alpha }
}\right) \left[ 2+ProductLog\left( 2e^{-1+\frac{2}{\alpha }}\right) \right] }
,
\end{equation}%
The horizon of the extremal black hole $r_{e}$ and critical value of spin
parameter $a_{c}$ verses $\alpha $ is plotted in FIG:\ref{reandac} (b) which
shows that size of the extremal black hole has minimum value for $\alpha
=2/3 $. It decreases for $0<\alpha <2/3$ and increase for $2/3<\alpha $. For 
$\tilde{\alpha}=1/(1+e)$, $a_{c}$ has minimum value $\tilde{a}_{c}\approx
0.855 $. Further, for $0<\alpha <\tilde{\alpha}$, $a_{c}$ decreases while $%
1/(1+e)<\alpha $ it increases.

We have plotted $y=r^{2}-2r+a^{2}$ for different values of $a$ and $%
y=-\alpha r\ln \left( \frac{r}{|\alpha |}\right) $ for negative $\alpha $ in
FIG:\ref{Fig4}(a)-(c) and for positive $\alpha$ in FIG:\ref{Fig4}(d)-(f). In
each case values of $r$ for which these curves intersect are horizons of the
black hole. It is seen that for any value of $-2\leq \alpha $, if $a<a_{c}$
the curves intersect for two values of $r$ that are locations of inner
horizon ($r_{-}$) and event horizon ($r_{+}$). If $a=a_{c}$ the horizons
merge into a single horizon $r_{e}$ the horizon of extremal black hole and
if $a>a_{c}$ the curve does not intersect each other that is no solution of
horizon equation. Thus, we conclude that for any fixed $-2\leq \alpha $ the
line element \eqref{LE} represents a black hole with two horizons $\ r_{-}$
and $r_{+}$ only if $a<a_{c}$. For $a=a_{c}$, the two horizons $r_{+}$ and $%
r_{-}$ merge into a single horizon $r_{e}$ and \eqref{LE} is a extremal
black hole. However, for any $a>a_{c}$, the line element is a naked
singularity.

\section{Spin Precession Frequency}

In this section, we will discuss the spin precession frequency of a test
gyroscope attached to a stationary observer with respect to some fixed star
due to the frame dragging effects of the Kerr-like black hole in the PFDM. The precession frequency ($\Omega _{p}$) of a test
gyroscope attached to a stationary observer having $4-$ velocity $%
u=(-K^{2})^{-1/2}K$ in a stationary spacetime with timelike Killing vector
field $K=\partial_{0}+\Omega \partial _{t}$ is defined by \cite{NS}

\begin{equation}
\vec{\Omega}_{p}=\frac{\varepsilon _{ckl}}{2\sqrt{-g}\left( 1+2\Omega \frac{
g_{0c}}{g_{00}}+\Omega ^{2}\frac{g_{cc}}{g_{00}}\right) }\left[ \left(
g_{0c,k}-\frac{g_{0c}}{g_{00}}g_{00,k}\right) +\Omega \left( g_{cc,k}-\frac{
g_{cc}}{g_{00}}g_{00,k}\right) +\Omega ^{2}\left( \frac{g_{0c}}{g_{00}}
g_{cc,k}-\frac{g_{cc}}{g_{00}}g_{0c,k}\right) \right] \partial _{l},
\label{SPD}
\end{equation}%
where $\varepsilon _{ckl}$ is the Levi-Civita symbols and $g$ is the
determinant of the metric $g_{\mu \nu }$. Using the metric coefficients from %
\eqref{LE} in \eqref{SPD} yields

\begin{equation}
\vec{\Omega}_{p}=\frac{\left( F\sqrt{\Delta }\cos \theta \right) \hat{r}%
\text{ }+\left( H\sin \theta \right) \hat{\theta}\text{ }}{\Sigma ^{3/2}%
\left[ \Sigma -\left\{ 2Mr-\alpha r\ln \left( \frac{r}{|\alpha |}\right)
\right\} \left( 1-2\Omega a\sin ^{2}\theta \right) -\Omega ^{2}\sin
^{2}\theta \left\{ \left( r^{2}+a^{2}\right) \Sigma +a^{2}\sin ^{2}\theta
\left( 2Mr-\alpha r\ln \left( \frac{r}{|\alpha |}\right) \right) \right\} %
\right] },  \label{Op}
\end{equation}%
where 
\begin{eqnarray}
F &=&a\left\{ 2Mr-\alpha r\ln \left( \frac{r}{|\alpha |}\right) \right\} -%
\frac{\Omega }{8}\left\{ 3a^{4}+8r^{4}+8a^{2}r\left( 2M+r\right)
+a^{2}\left( a^{2}\cos 4\theta -8\alpha r\ln \left( \frac{r}{|\alpha |}%
\right) +4\cos 2\theta \left( 2\Delta -a^{2}\right) \right) \right\}  \notag
\\
&&+\Omega ^{2}a^{3}\left\{ 2Mr-\alpha r\ln \left( \frac{r}{|\alpha |}\right)
\right\} \sin ^{4}\theta ,  \label{F} \\
H &=&a\left[ M\left( r^{2}-a^{2}\cos ^{2}\theta \right) +\frac{\alpha }{2}%
\left\{ \Sigma -\left( r^{2}-a^{2}\cos ^{2}\theta \right) \ln \left( \frac{r%
}{|\alpha |}\right) \right\} \right]  \notag \\
&&+\Omega \left[ a^{4}r\cos ^{4}\theta +r^{2}\left( r^{3}-a^{2}M\left(
1+\sin ^{2}\theta \right) -3Mr^{2}\right) +a^{2}\cos ^{2}\theta \left\{
2r^{3}+a^{2}M\left( 1+\sin ^{2}\theta \right) -Mr^{2}\right\} \right.  \notag
\\
&&\left. -\frac{\alpha }{16}\left\{ a^{2}\left( 5a^{2}+16r^{2}\right)
+8r^{4}+\left( 5a^{4}-16a^{2}r^{2}-24r^{4}\right) \ln \left( \frac{r}{%
|\alpha |}\right) +a^{4}\left( 4\cos 2\theta -\cos 4\theta \right) \left\{
1+\ln \left( \frac{r}{|\alpha |}\right) \right\} \right\} \right]  \notag \\
&&+a\Omega ^{2}\sin ^{2}\theta \left[ M\left\{ r^{2}\left(
3r^{2}+a^{2}\right) +a^{2}\cos ^{2}\theta \left( r^{2}-a^{2}\right) \right\}
+\frac{\alpha }{2}\left\{ a^{2}\cos ^{2}\theta \left\{ r^{2}+a^{2}+\left(
a^{2}-r^{2}\right) \ln \left( \frac{r}{|\alpha |}\right) \right\} \right.
\right.  \notag \\
&&\left. \left. +r^{2}\left\{ r^{2}+a^{2}-\left( 3r^{2}+a^{2}\right) \ln
\left( \frac{r}{|\alpha |}\right) \right\} \right\} \right] ,  \label{G}
\end{eqnarray}%
and $\hat{r}$, $\hat{\theta}$ are the unit vectors in $r$ and $\theta $
directions, respectively. In the limiting case $\alpha =0$, the spin
precession of a Kerr black hole is successfully obtained \cite{NS}. Note
that the above expression of the precession frequency \eqref{SPD} is valid
only for a timelike observer at fixed $r$ and $\theta $ which gives the
restriction on the angular velocity $\Omega $ of the observer 
\begin{equation}
\Omega _{-}(r,\theta )<\Omega (r,\theta )<\Omega _{+}\left( r,\theta \right)
,  \label{range}
\end{equation}%
with 
\begin{equation}
\Omega _{\pm }=\frac{a\sin \theta \left\{ 2Mr-\alpha r\ln \left( \frac{r}{%
|\alpha |}\right) \right\} \pm \Sigma \sqrt{\Delta }}{\sin \theta \left[
\left( r^{2}+a^{2}\right) \Sigma +a^{2}\sin ^{2}\theta \left\{ 2Mr-\alpha
r\ln \left( \frac{r}{|\alpha |}\right) \right\} \right] }.  \label{omegapm}
\end{equation}%
At the black hole horizons, $\Omega _{+ }$ and $\Omega _{-}$ coincide and no timelike
observer can exist there and hence the expression for precision frequency $%
\Omega _{p}$ is not valid at the horizons but still we can study the
behavior of precession frequency near the black hole horizon.

\subsection{Lense-Thirring precession frequency}

The expression of the precession frequency \eqref{Op} is valid for all the
stationary observers inside and outside the ergosphere if their angular
velocity $\Omega$ is in the restricted range given by \eqref{range}. The
precession frequency contains the effects because of the spacetime rotation
(LT precession) as well as due to spacetime curvature (geodetic precession).
If we set $\Omega =0$ in \eqref{Op}, the expression for LT precession
frequency for Kerr-Like black hole in PFDM is obtained
as

\begin{figure}[!ht]
\centering
%\captionsetup{justification=centering} 
\minipage{0.31\textwidth} %
\includegraphics[width=2.2in,height=1.6in]{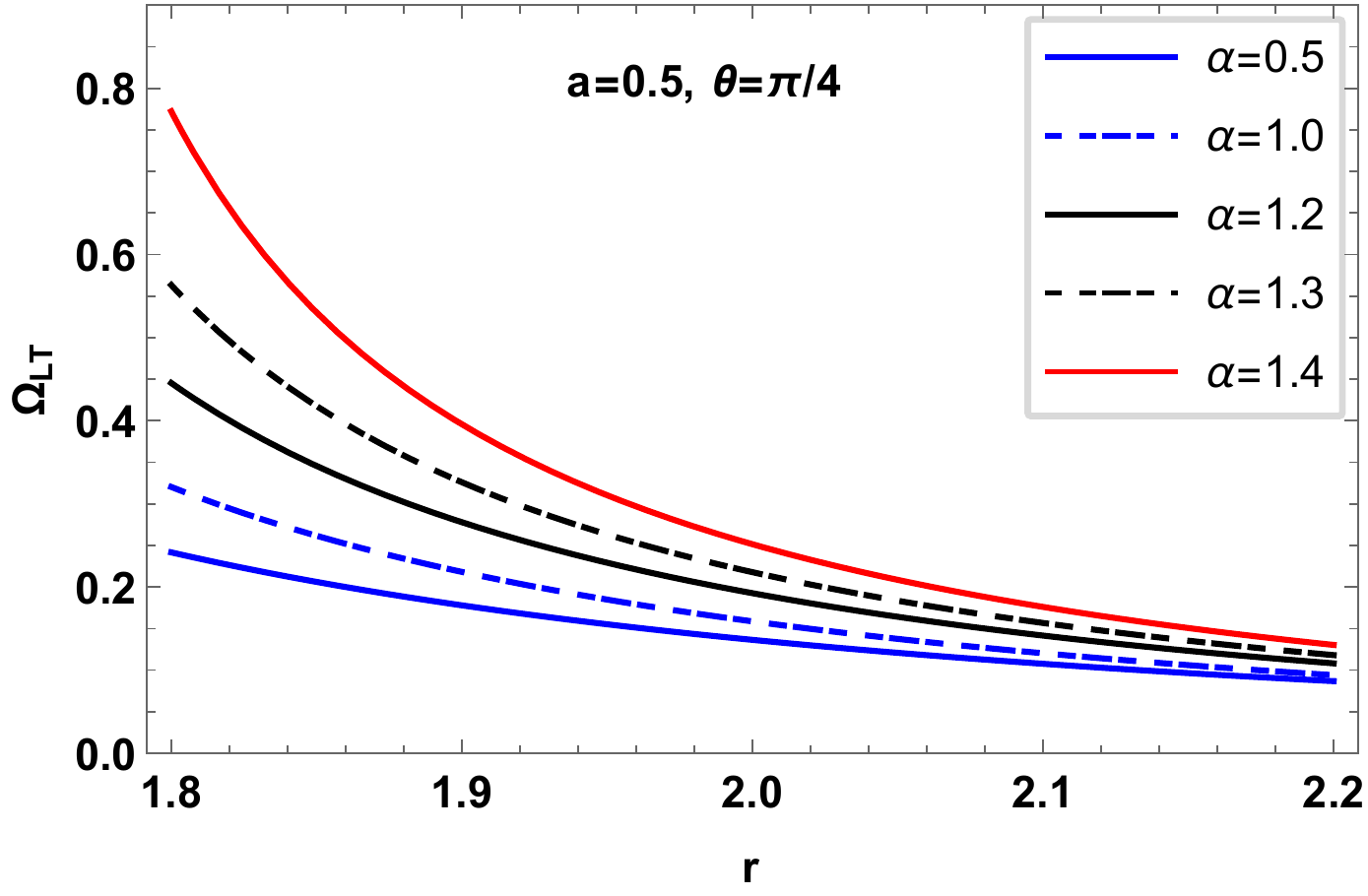}\newline
(a)  \endminipage\hfill \minipage{0.31\textwidth} %
\includegraphics[width=2.2in,height=1.6in]{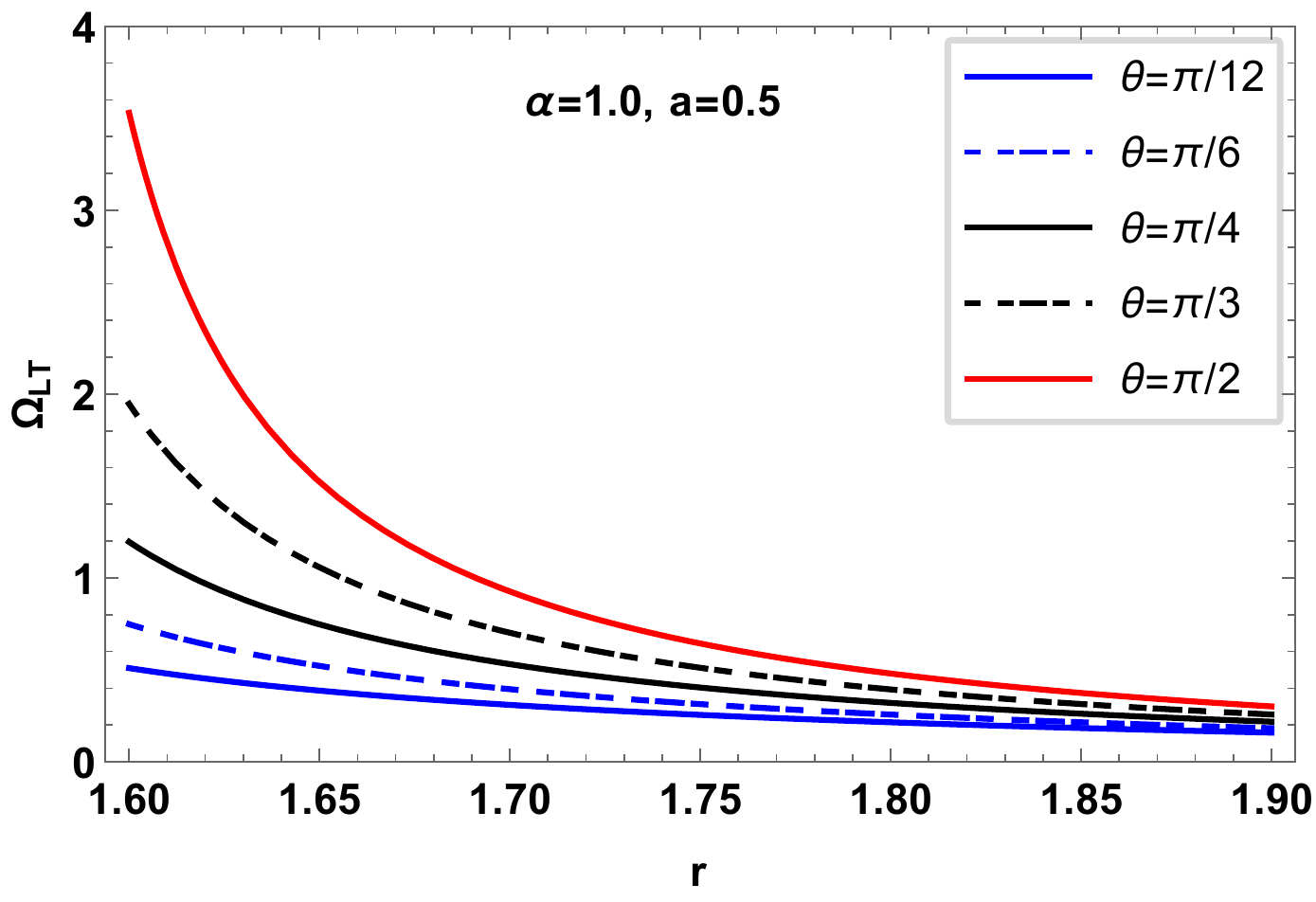}\newline
(b) %
\par
\endminipage\hfill \minipage{0.31\textwidth} %
\includegraphics[width=2.2in,height=1.6in]{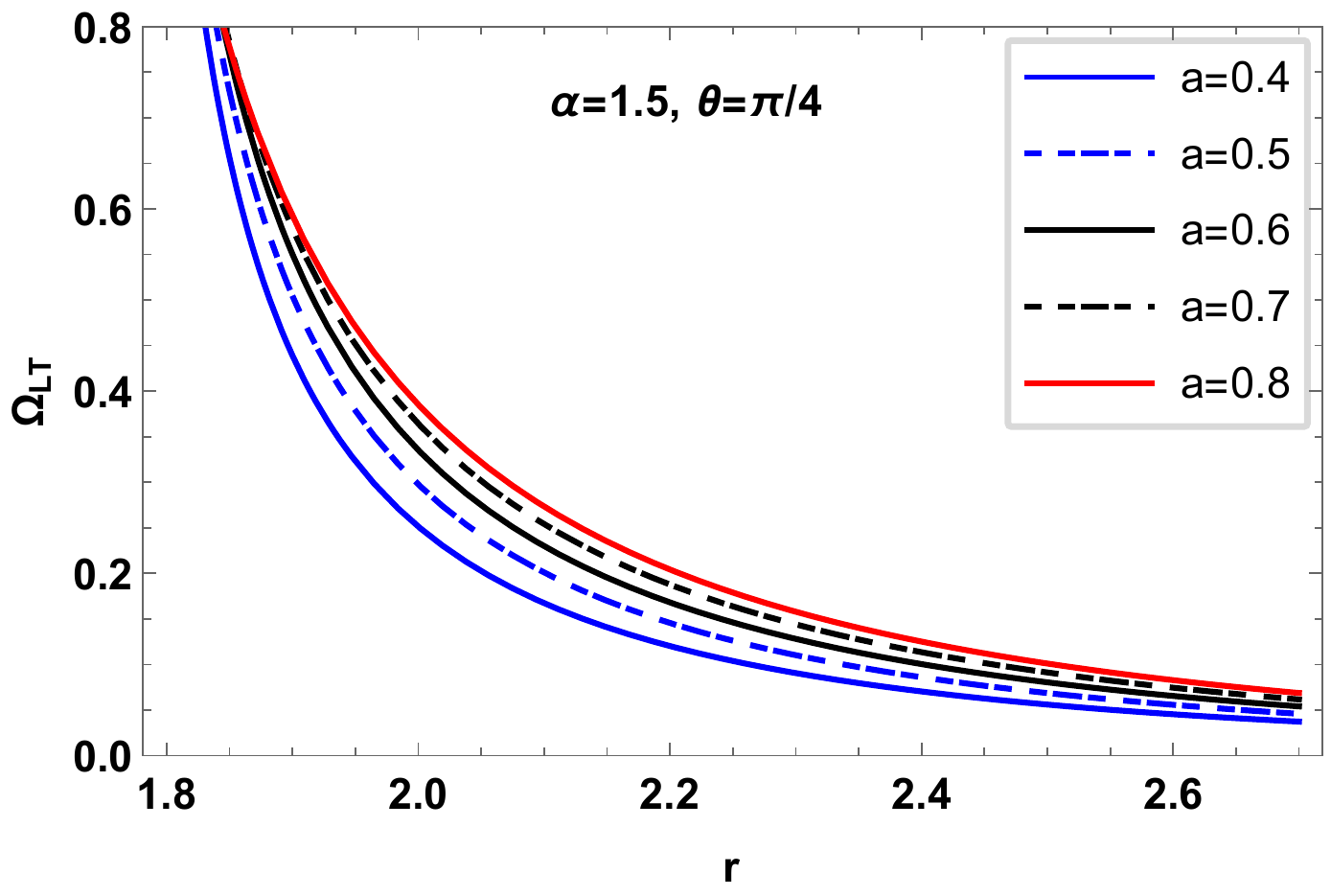}\newline
(c) %
\par
\endminipage\hfill
\caption{{\protect\footnotesize The LT precession frequency $\Omega_{LT}$
(in $M^{-1}$ ) verse $r$ (in $M$) for different parameters is plotted.}}
\label{LTFig}
\end{figure}

\begin{equation}  \label{LTvf}
\vec{\Omega}_{LT}=a\frac{\left[ 2Mr-\alpha r\ln \left( \frac{r}{|\alpha |}
\right) \right] \sqrt{\Delta }\cos \theta \hat{r}\text{ }+\sin \theta \left[
M\left( r^{2}-a^{2}\cos ^{2}\theta \right) +\frac{\alpha }{2}\left\{ \Sigma
-\left( r^{2}-a^{2}\cos ^{2}\theta \right) \ln \left( \frac{r}{|\alpha |}
\right) \right\} \right] \hat{\theta}\text{ }}{\Sigma ^{3/2}\left[ \Sigma
-\left\{ 2Mr-\alpha r\ln \left( \frac{r}{\alpha }\right) \right\} \right] }.
\end{equation}

\begin{figure}[!ht]
\centering
\minipage{0.30\textwidth} %
\includegraphics[width=2.0in,height=2.0in]{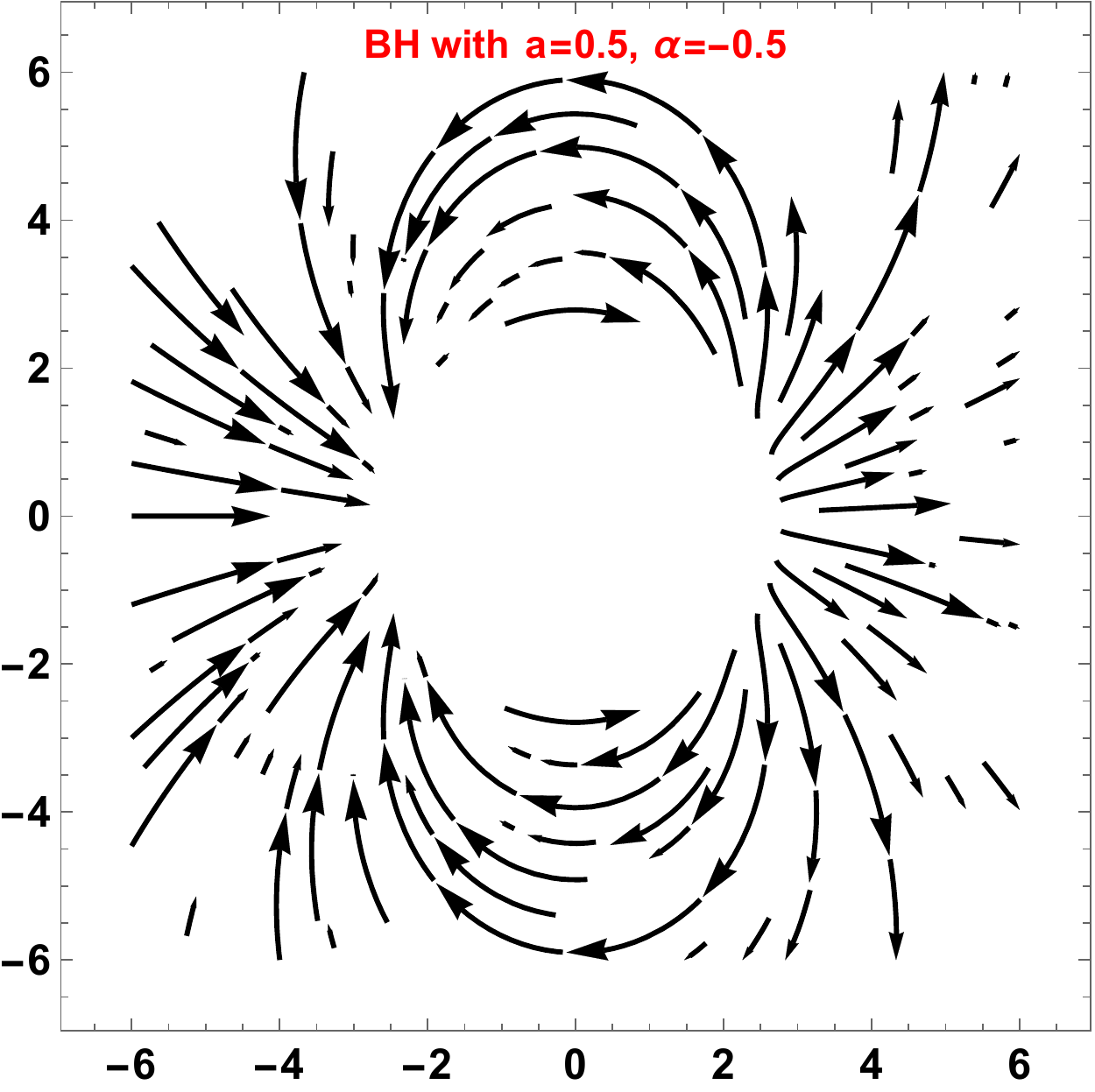}\newline
(a)  \label{7} \endminipage\hfill \minipage{0.30%
\textwidth} \includegraphics[width=2.0in,height=2.0in]{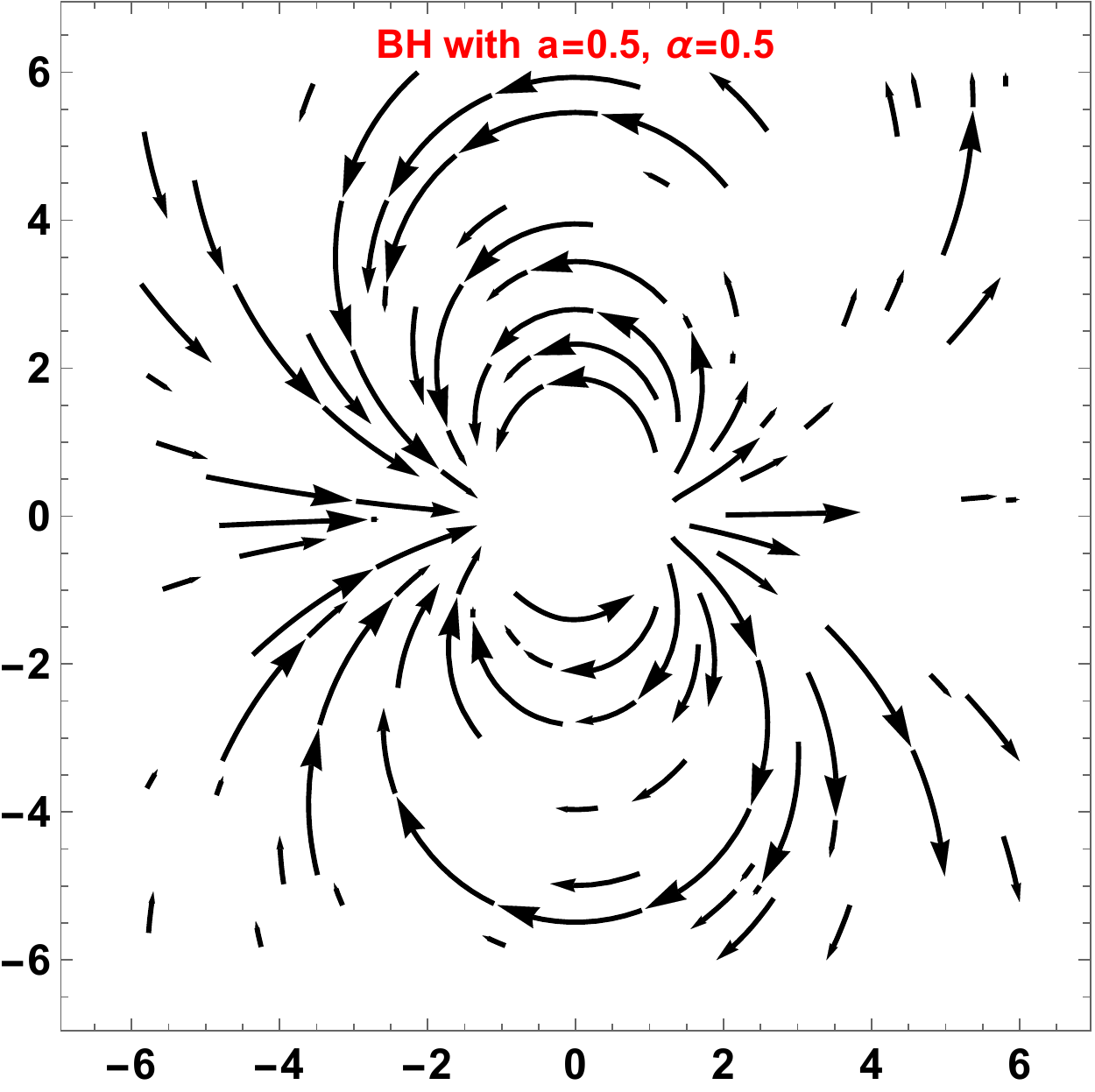}%
\newline
(b)  \label{7} \endminipage\hfill \minipage{0.30%
\textwidth} \includegraphics[width=2.0in,height=2.0in]{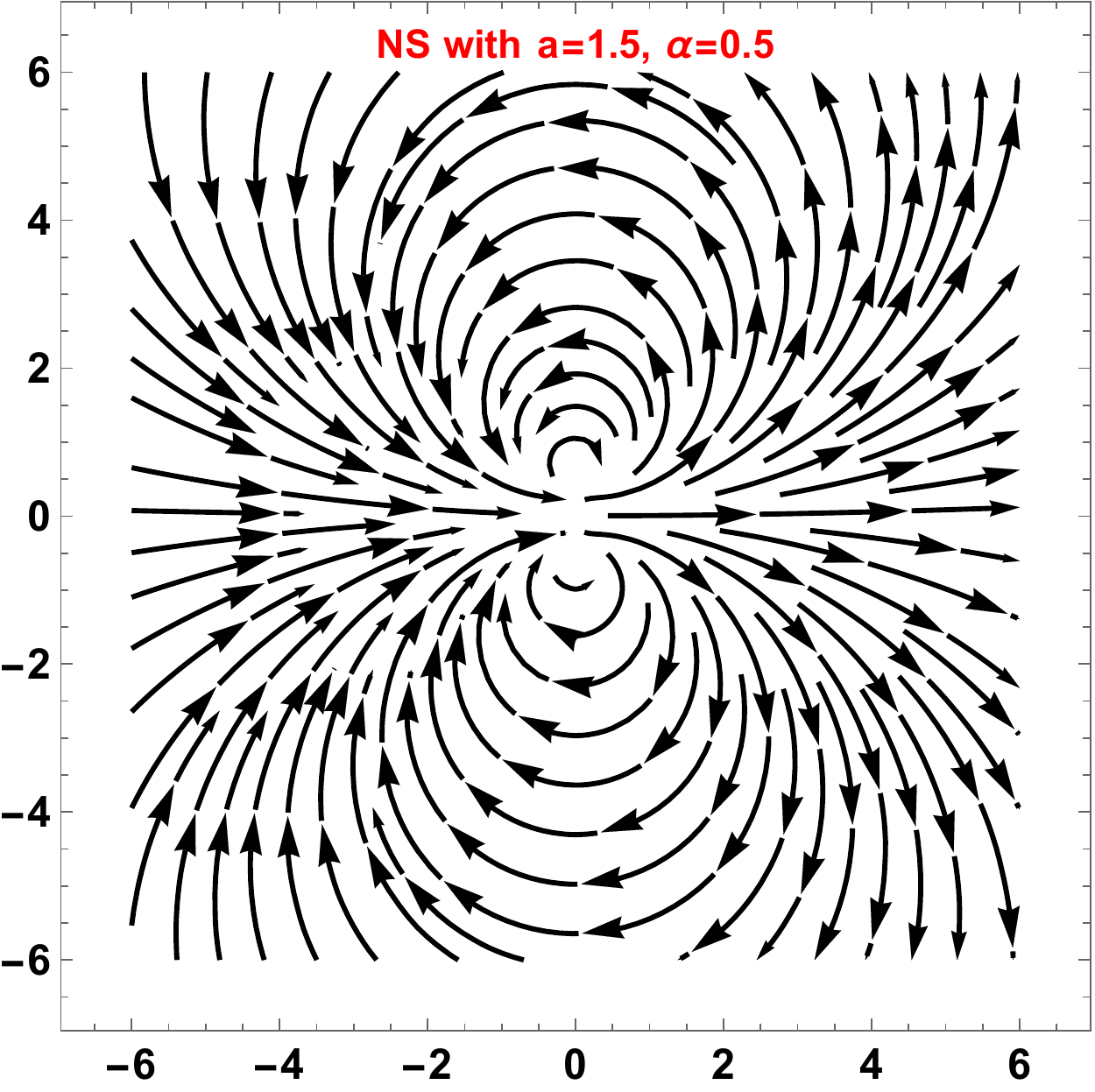}%
\newline
(c)  \endminipage\hfill
\caption{{\protect\footnotesize The vector field of the LT-~precession
frequency \eqref{LTvf} (in Cartesian plane corresponding to $(r,\protect%
\theta)$) for black holes is plotted in panels (a) and (b) for negative and positive
 $\protect\alpha$ and for naked singularity in panel (c). The field
lines show that for black hole the vector field is defined outside the
ergoshpere only, while for naked singularities it is finite up to the ring
singularity along all the directions.}}
\label{VF}
\end{figure}
The magnitude of the LT precession frequency is given by
\begin{equation}  \label{LT}
\Omega _{LT}=a\frac{\sqrt{\left[ 2Mr-\alpha r\ln \left( \frac{r}{|\alpha |}
\right) \right] ^{2}|\Delta |\cos ^{2}\theta \text{ }+\left[ M\left(
r^{2}-a^{2}\cos ^{2}\theta \right) +\frac{\alpha }{2}\left\{ \Sigma -\left(
r^{2}-a^{2}\cos ^{2}\theta \right) \ln \left( \frac{r}{|\alpha |}\right)
\right\} \right] ^{2}\sin ^{2}\theta }\text{ }}{\Sigma ^{3/2}\left\vert
\Sigma -\left\{ 2Mr-\alpha r\ln \left( \frac{r}{|\alpha |}\right) \right\}
\right\vert }.
\end{equation}

The magnitude of LT precession frequency $\Omega_{LT}$ is plotted against $r$
in the FIG: (\ref{LTFig}), which indicates that the LT precession frequency
increases with increasing the rotation $a$ of the black hole as well as the
dark energy parameter $\alpha$. Further, for $\alpha>0$ the LT precession
frequency is minimum near polar axis ($\theta=0$) and increases towards the
equatorial plane ($\theta=\pi/2$) whereas for $\alpha<0$ it is minimum in
equatorial plane and increases towards the polar axis. The vector field of
the LT-precession frequency for the black hole and naked singularity in FIG:(%
\ref{VF}) which shows that LT precession frequency for the black hole remains
finite outside the ergoregion and will diverge at ergoregion while for
naked singularity it remains finite up to the ring singularity.

\section{Geodetic precession}

For $a=0$, the line element \eqref{LE} reduces to the Schwarzschild black
hole in PFDM \cite{SCindark}. The Schwarzschild black
hole in PFDM is non
rotating and have zero precession due to the frame dragging effects.
However, due to spacetime curvature the precession frequency $\Omega_{p}$ of
a test gyroscope is nonzero which is because of curvature of spacetime and
called geodetic precession frequency. The geodetic precession effects can be
obtained as

\begin{equation}
\vec{\Omega}_{p}|_{a=0}=\Omega \frac{\left( -\cos \theta \sqrt{
r^{2}-2Mr+\alpha r\left( \frac{r}{|\alpha |}\right) }\right) \hat{r}+\sin
\theta \left( r-3M-\frac{\alpha }{2}\left\{ 1-3\ln \left( \frac{r}{|\alpha |}
\right) \right\} \right) \hat{\theta}}{r-2M+\alpha \ln \left( \frac{r}{
|\alpha |}\right) -\Omega ^{2}r^{3}\sin ^{2}\theta }.
\end{equation}%
Due to spherically symmetric geometry of the static black hole in the PFDM the geodetic precession frequency is same over any
spherical symmetric surface around the black hole. Thus, without loss of
generality we can set $\theta=\pi/2$ and study the geodetic frequency in
equatorial plane. In this plane for any observer in circular orbit the
magnitude of precession frequency is equal to the Kepler frequency given by 
\begin{equation}
{\Omega}_{p}|_{a=0}\equiv\Omega _{\text{Kep}}=\left[ \frac{M}{r^{3}}+\frac{\alpha 
}{2r^{3}}\left\{ 1-\ln \left( \frac{r}{|\alpha |}\right) \right\} \right]
^{1/2}.
\end{equation}

\begin{figure}[!ht]
\centering
%\captionsetup{justification=centering}
 \minipage{0.50\textwidth} %
\includegraphics[width=8.2cm,height=5.4cm]{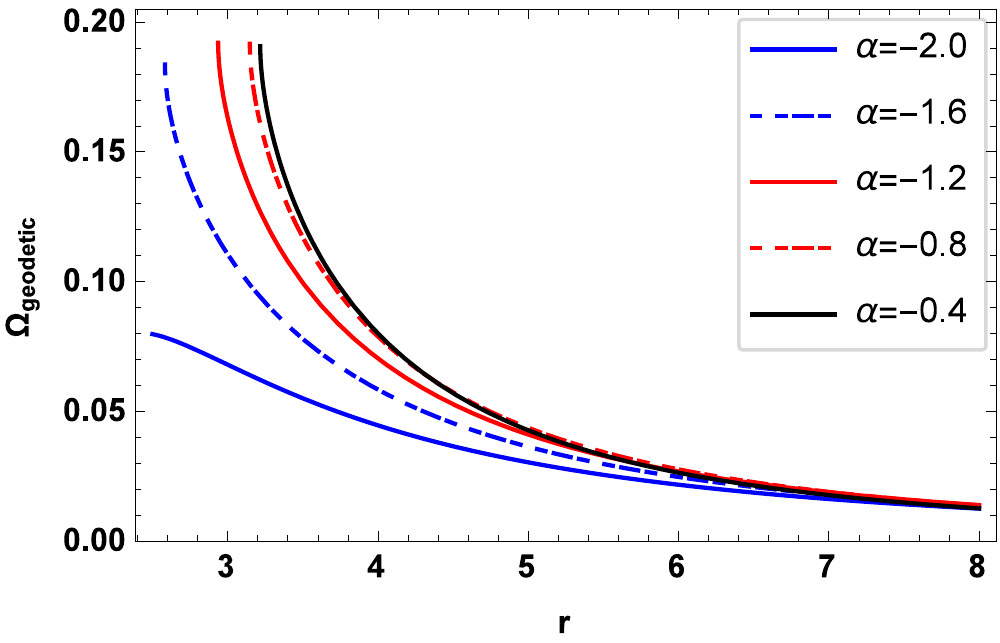}\newline
(a) \label{rea} \endminipage\hfill \minipage{0.50\textwidth} %
\includegraphics[width=8.0cm,height=5.6cm]{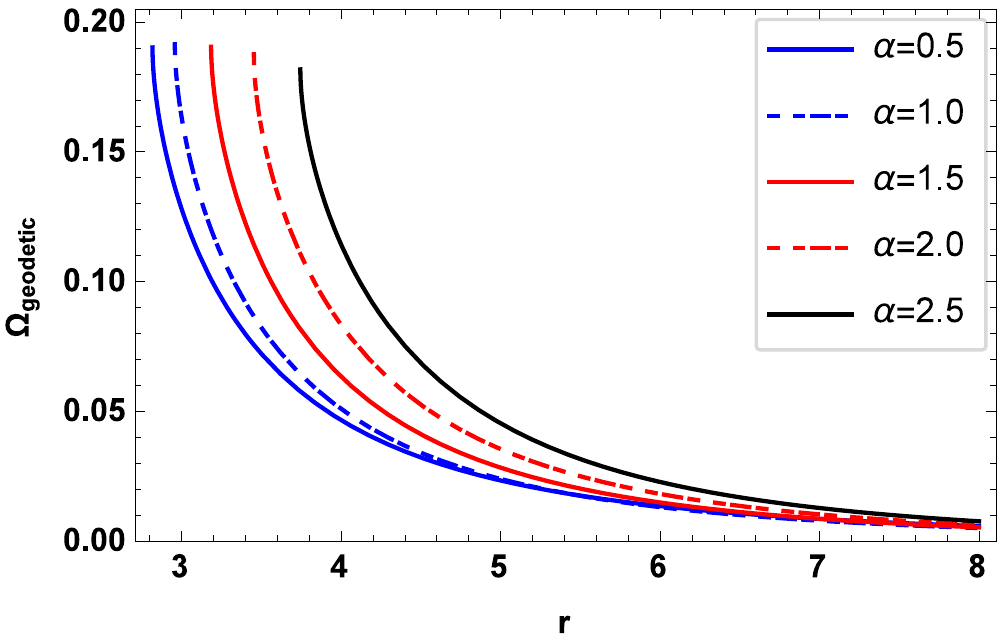}\newline
(b) \label{reb} \endminipage\hfill
\caption{{\protect\footnotesize The geodetic precession frequency $\Omega_{%
\text{geodetic}}$ verse $r$ for $\protect\alpha<0$ in panel (a) and for $%
\protect\alpha> 0$ in panel (b) is plotted which shows that the geodetic
precession frequency increases with increasing $\protect\alpha$. }}
\label{geo}
\end{figure}

The above expression for precession frequency is valid for Copernican frame a frame that does not rotate relative to the inertial frame at asymptotic infinity i.e. the fixed stars,
computed with respect to proper time $\tau$ which is related with the coordinate
time $t$ via

\begin{equation}
d\tau =\sqrt{1-\frac{3M}{r}-\frac{\alpha }{2r}\left\{ 1-\ln \left( \frac{r}{
|\alpha |}\right) \right\} }dt.
\end{equation}%
Using this transformation the geodetic precession frequency associated with
the change in the angle of the spin vector after one complete revolution of
the observer around a black hole in the coordinate basis is given by 
\begin{equation}
\Omega _{\text{geodetic}}=\left[ \frac{M}{r^{3}}+\frac{\alpha }{2r^{3}}
\left\{ 1-\ln \left( \frac{r}{|\alpha |}\right) \right\} \right]
^{1/2}\left( 1-\sqrt{1-\frac{3M}{r}-\frac{\alpha }{2r}\left\{ 1-\ln \left( 
\frac{r}{|\alpha |}\right) \right\} }\right).
\end{equation}
In this absence of PFDM around the black hole that is
for $\alpha=0$, the geodetic precession frequency for a Schwarzschild black
hole successfully recovered \cite{geosc1,geosc2}. The geodetic precession
frequency is plotted against $r$ for negative and positive $\alpha$ in panel 
$(a)$ and $(b)$ of FIG: (3), which shows that with increasing PFDM parameter $\alpha$ the magnitude of the geodetic precession
frequency increases. Further, for any fixed $\alpha$ the geodetic precession
frequency decreases with increasing the radius of the circular orbit. 
\begin{figure}[!ht]
\centering
\minipage{0.50\textwidth} %
\includegraphics[width=8.3cm,height=5.8cm]{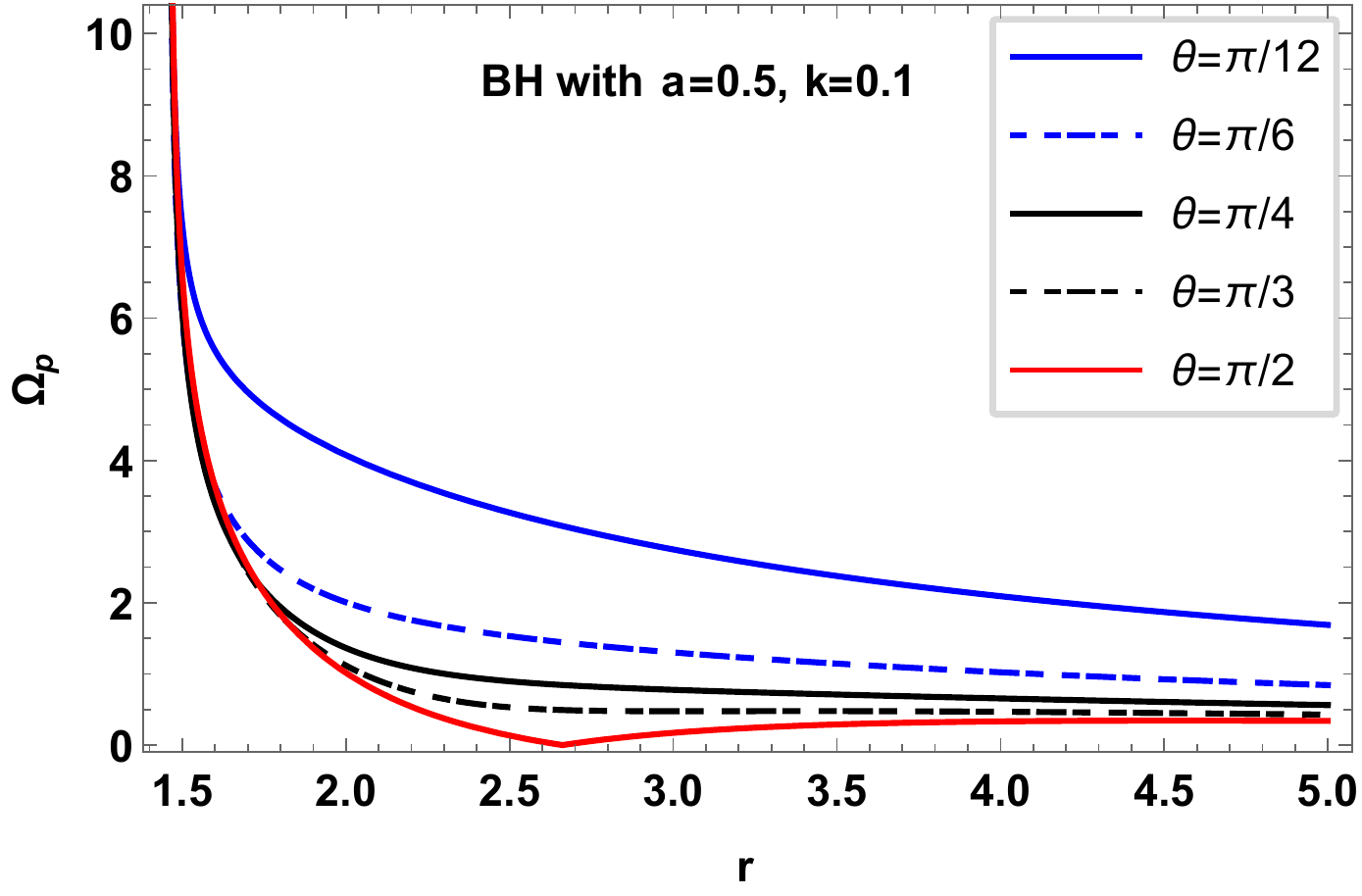}\newline
(a)   \label{6} \endminipage%
\hfill \minipage{0.50\textwidth} %
\includegraphics[width=8.3cm,height=5.8cm]{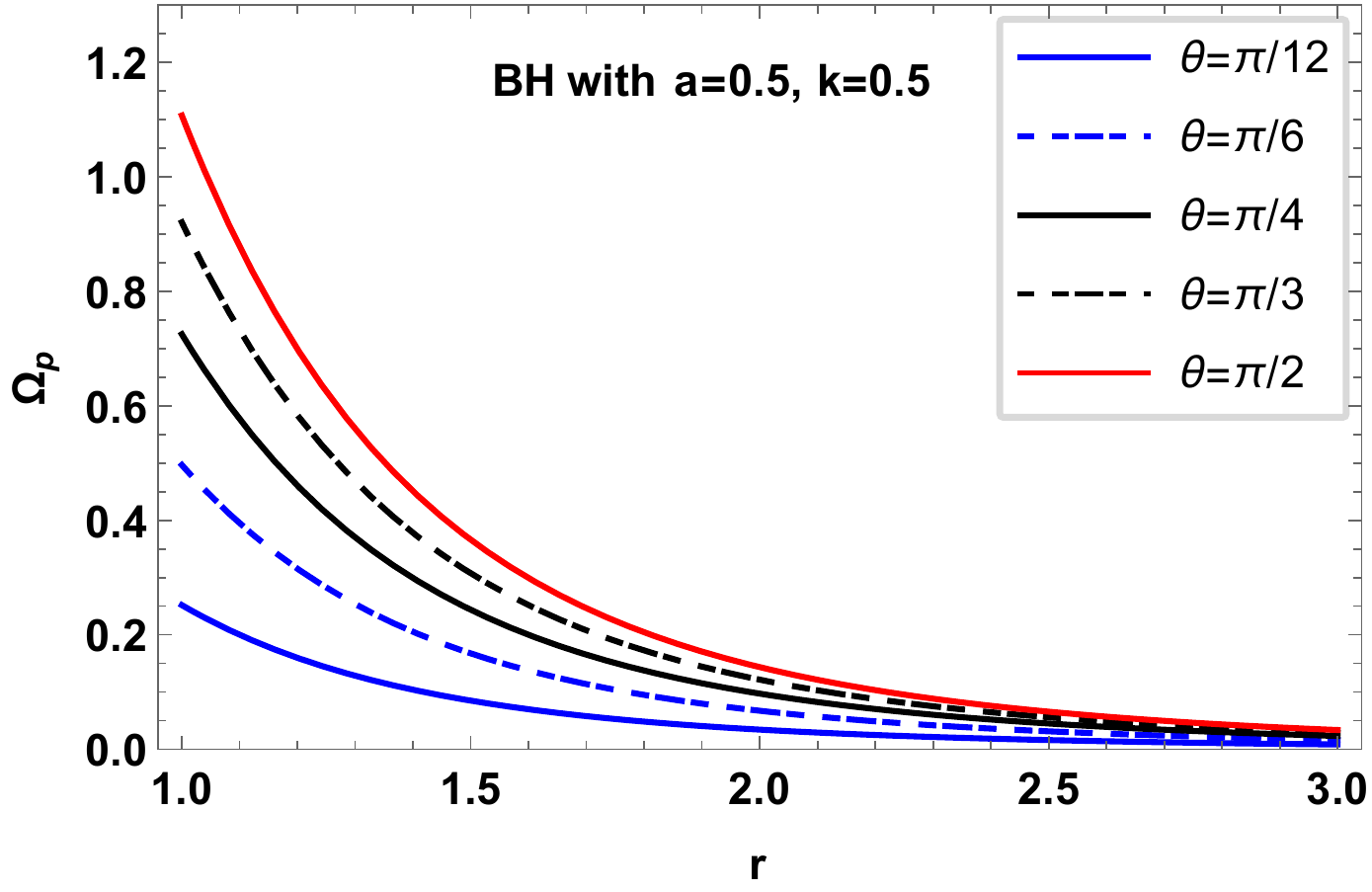}\newline
(b)  \label{7} \endminipage%
\hfill \minipage{0.50\textwidth} %
\includegraphics[width=8.3cm,height=5.8cm]{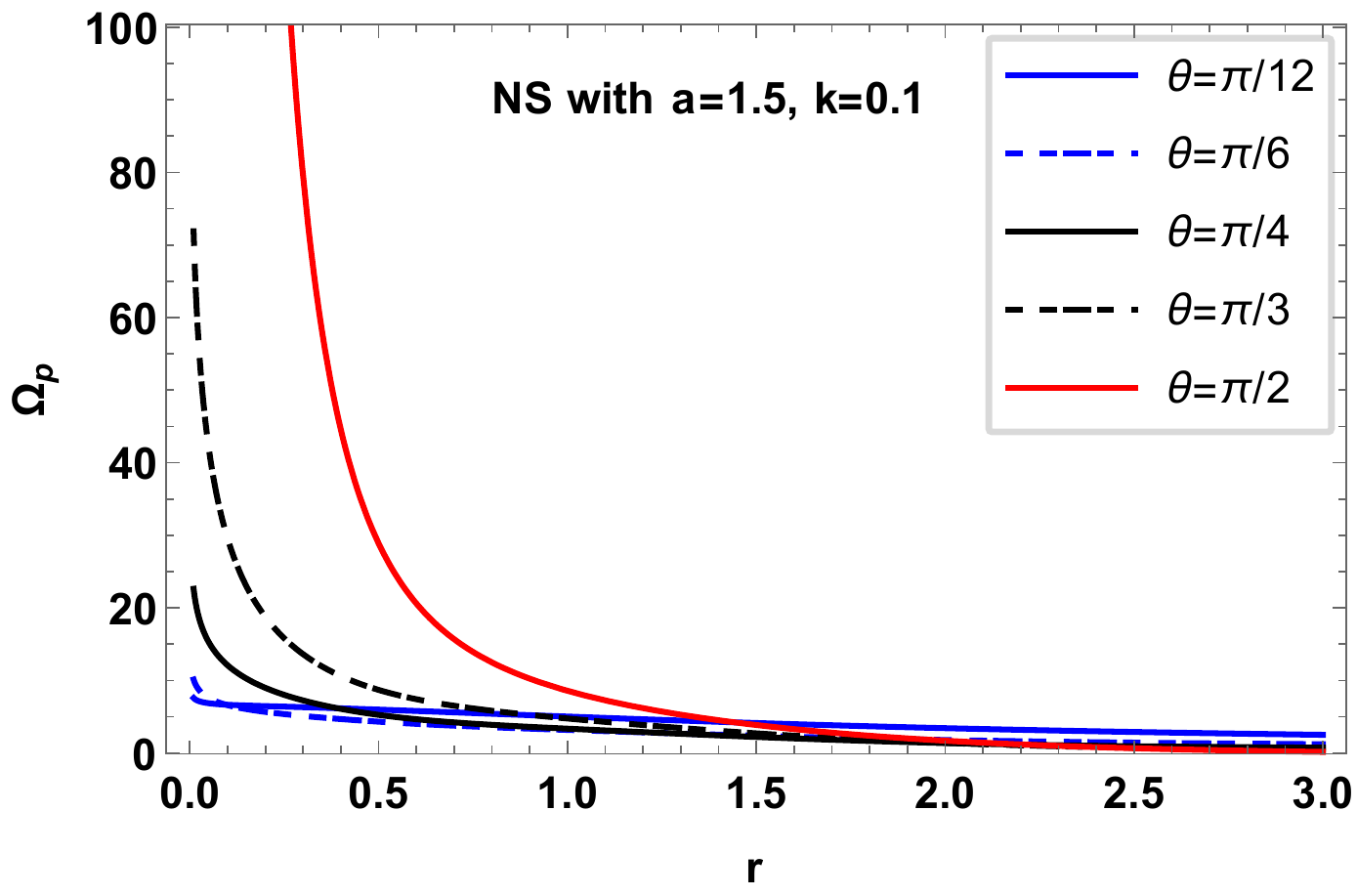}\newline
(c) \label{6} \endminipage%
\hfill \minipage{0.50\textwidth} %
\includegraphics[width=8.3cm,height=5.8cm]{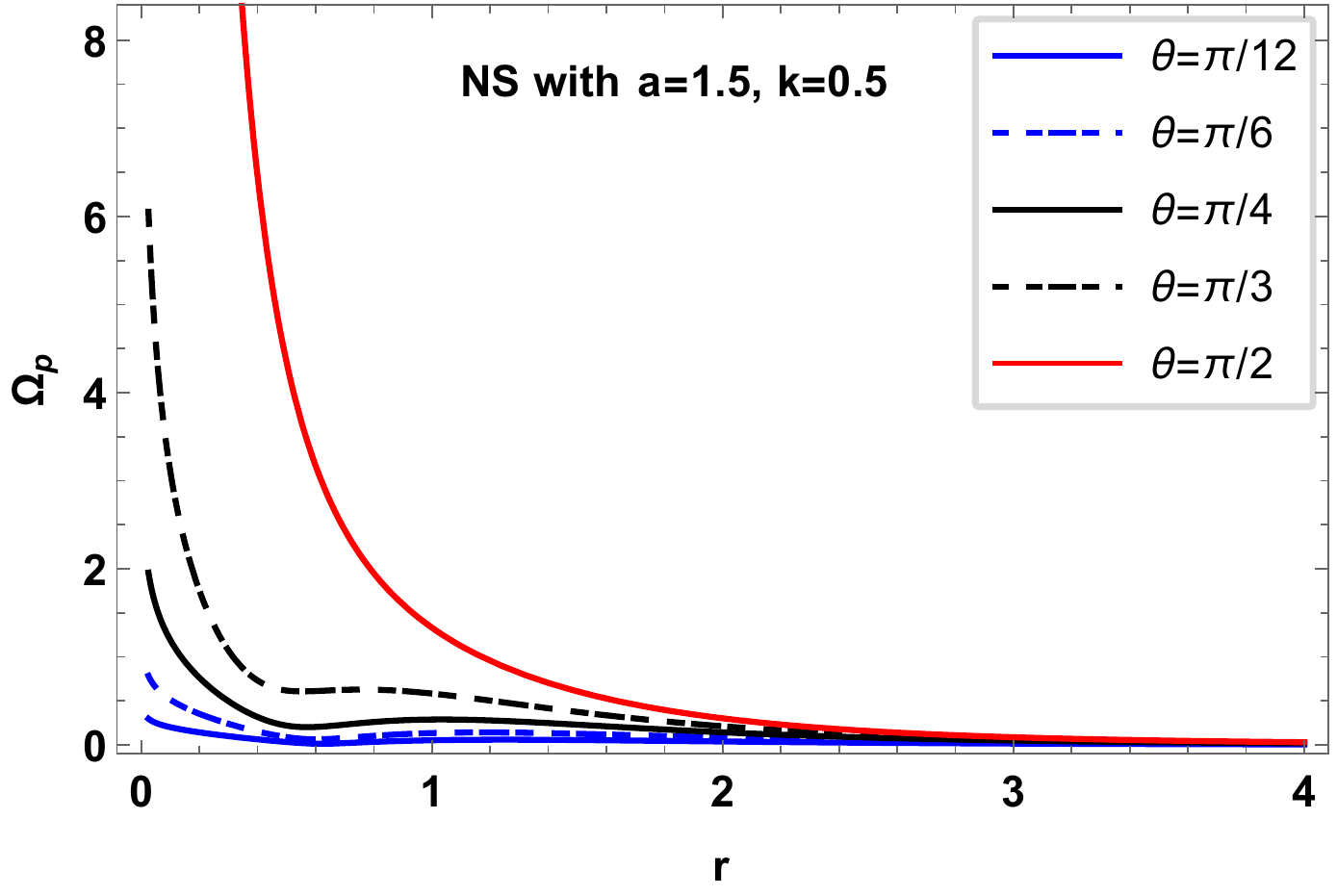}\newline
(d) \label{6} %
\endminipage\hfill \minipage{0.50\textwidth} %
\includegraphics[width=8.3cm,height=5.8cm]{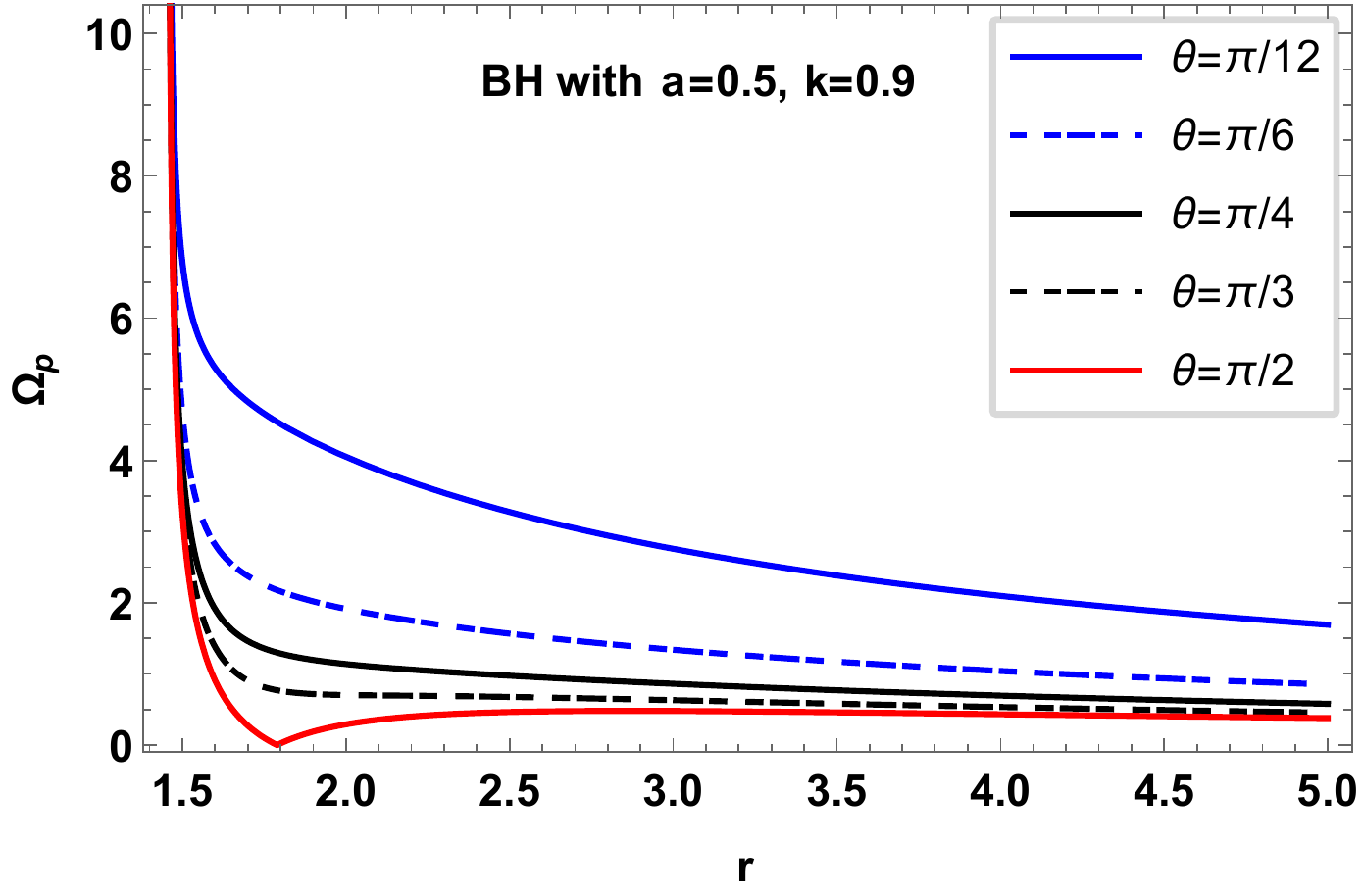}\newline
(e)  \label{6} \endminipage%
\hfill \minipage{0.50\textwidth} %
\includegraphics[width=8.3cm,height=5.8cm]{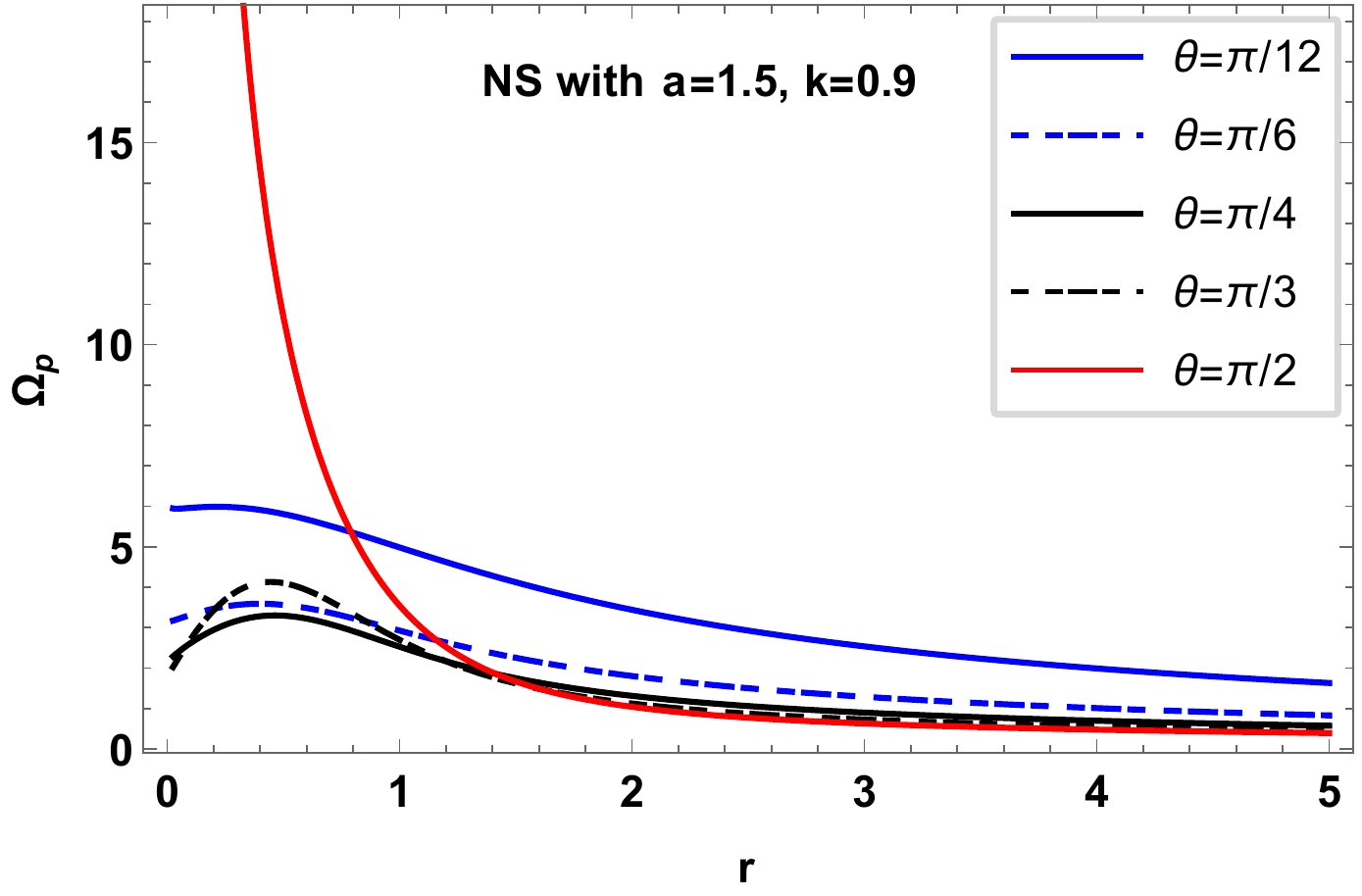}\newline
(f)  \label{7} %
\endminipage\hfill
\caption{{\protect\footnotesize We have plotted the magnitude of spin
precession frequency $\Omega_p$ (in $M^{-1}$ ) verse $r$ (in $M$) is plotted
for black hole in left column and for naked singularity in right column. For
black hole we take $a=0.5$, $\protect\alpha=1$ and for naked singularity we
take $a=1.5$ and $k=0.1,0.5,0.9$ in first, second and third row,
respectively. For black hole the precession frequency $\Omega_p$ diverges
for $k=0.1,0.9$ and remain finite for $k=0.5$ as the observer approaches the
event horizon along any direction, whereas for naked singularity case it
with remain finite along all direction except at ring singularity ($r=0$,~$%
\protect\theta=\protect\pi/2$).}}
\label{opfig}
\end{figure}

\section{Distinguishing black hole from naked singularity}

In this section, using the precession frequency of a gyroscope, we will
differentiate a Kerr-like black hole in PFDM from naked
singularity and verify our results as obtained in section II. For this we
first express the angular velocity of the timelike observer in term of a
parameter $k$ such that 
\begin{equation}
\Omega=k\Omega_{+}+(1-k)\Omega_{-},
\end{equation}
where $0<k<1$ and $\Omega_{\pm }$ given by \eqref{omegapm}. Thus for any
timelike observer the angular velocity defined as

\begin{equation}
\Omega =\frac{a\sin \theta \left\{ 2Mr-\alpha r\ln \left( \frac{r}{|\alpha |}
\right) \right\} -\left( 1-2k\right) \Sigma \sqrt{\Delta }}{\sin \theta %
\left[ \left( r^{2}+a^{2}\right) \Sigma +a^{2}\sin ^{2}\theta \left\{
2Mr-\alpha r\ln \left( \frac{r}{|\alpha |}\right) \right\} \right] },
\end{equation}
Note that an observer with angular velocity parameter $k=1/2$ is known as
zero-angular-momentum observer (ZAMO) and has an angular velocity 
\begin{equation}
\Omega=-\frac{g_{t\phi}}{g_{\phi\phi}}=\frac{a \left\{ 2Mr-\alpha r\ln
\left( \frac{r}{|\alpha |} \right) \right\} }{ \left( r^{2}+a^{2}\right)
\Sigma +a^{2}\sin ^{2}\theta \left\{ 2Mr-\alpha r\ln \left( \frac{r}{|\alpha
|}\right) \right\} },
\end{equation}
The gyroscope attached to ZAMO observer is locally nonrotating and useful to
study physical procession near astrophysical objects \cite{Bardeen}.
Further, the behavior of the precession frequency attached to ZAMO observer
is different from the all other observers this situation is explain in FIG:
(\ref{opfig}). Now the magnitude of the precession frequency $\Omega_{p}$ in term of
parameter $k$ is written as 
\begin{figure}[!ht]
\centering
%\captionsetup{justification=centering}
 \minipage{0.31\textwidth} %
\includegraphics[width=2.2in,height=1.6in]{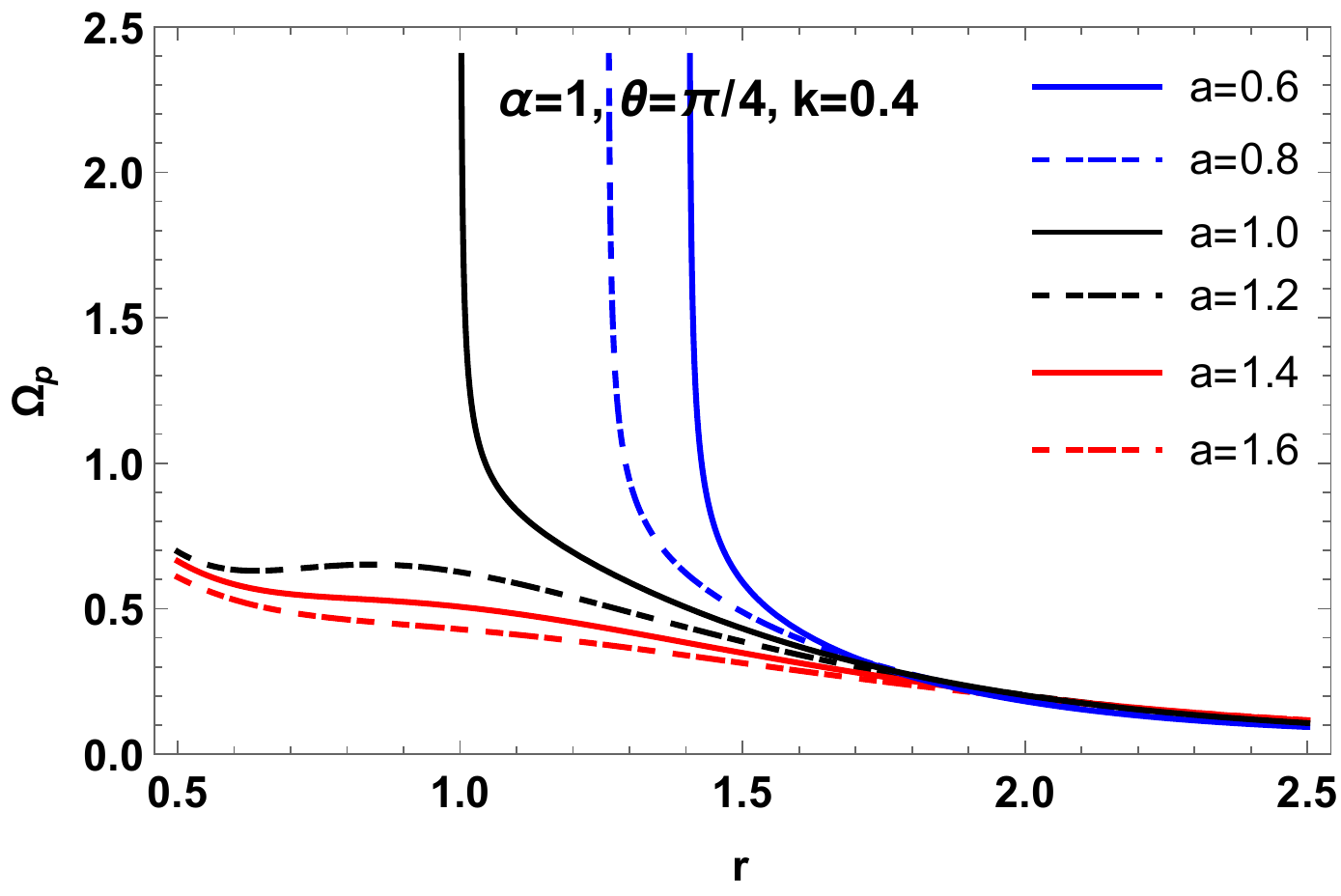}\newline
(a)  \label{6} \endminipage\hfill %
\minipage{0.31\textwidth} %
\includegraphics[width=2.2in,height=1.6in]{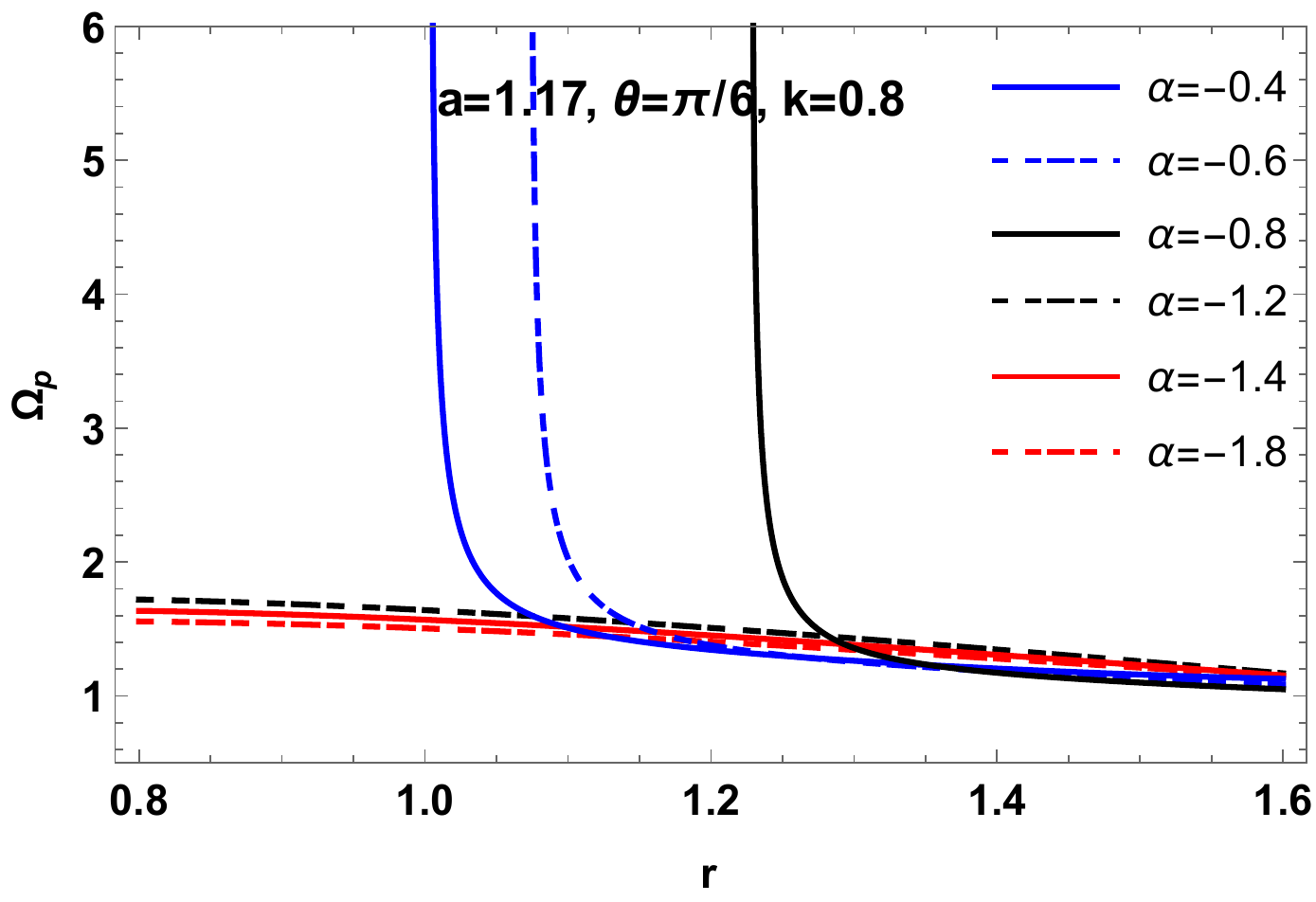}\newline
(b)  \label{7} \endminipage\hfill %
\minipage{0.31\textwidth} %
\includegraphics[width=2.2in,height=1.6in]{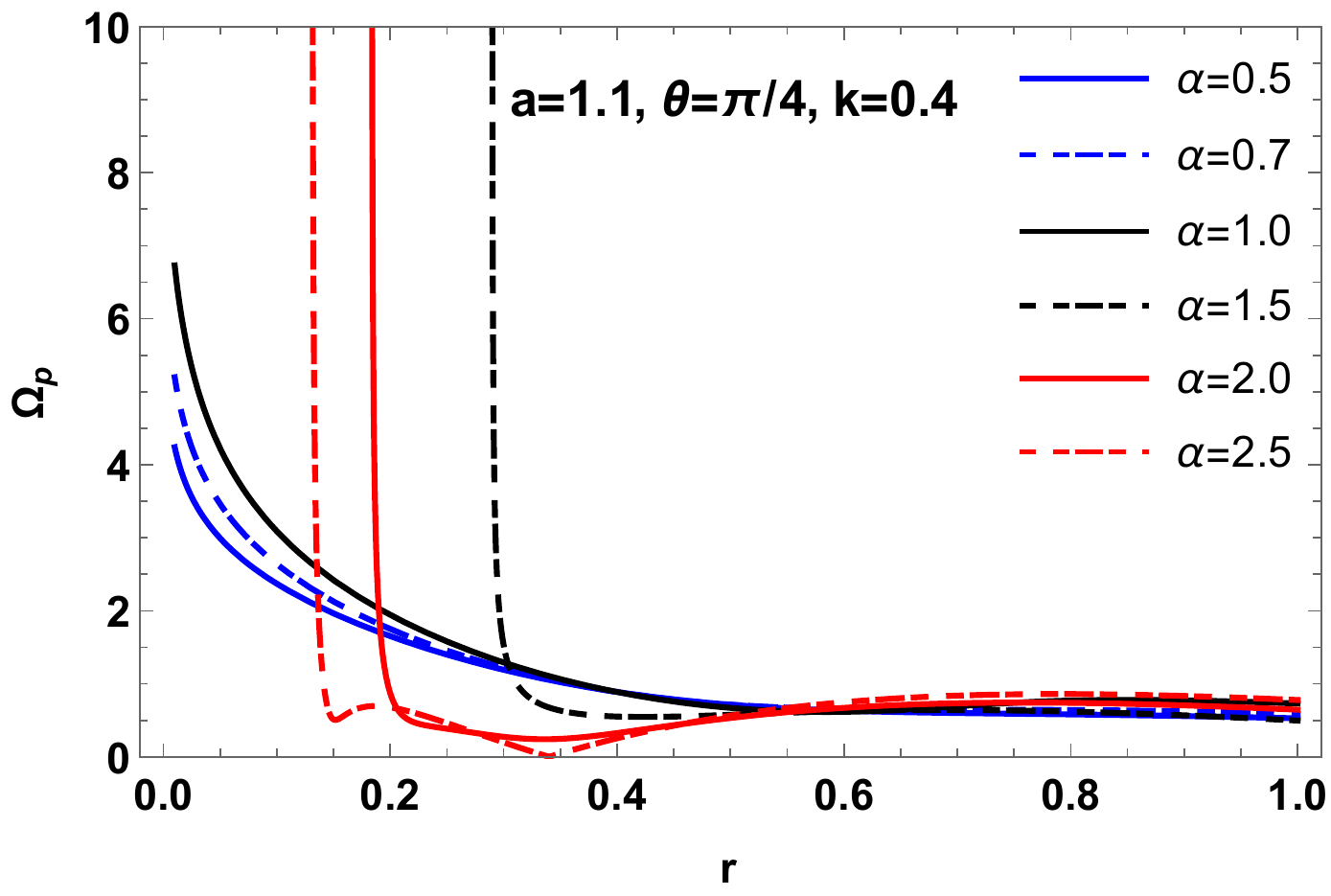}\newline
(c)  \label{7} \endminipage\hfill
\caption{{\protect\footnotesize The precession frequency $\Omega_p$ (in $%
M^{-1}$ ) verse $r$ (in $M$) for different parameters is plotted. The graphs
show that for black holes $\Omega_p$ diverges near the horizons while for
naked singularities it remain finite. }}
\label{opf}
\end{figure}
\begin{equation}
\Omega _{p}=\frac{\left\vert \left( r^{2}+a^{2}\right) \Sigma +a^{2}\sin
^{2}\theta \left( 2Mr-\alpha r\ln \left( \frac{r}{|\alpha |}\right) \right)
\right\vert }{4\Sigma ^{7/2}k\left( 1-k\right) \left\vert \Delta \right\vert 
}\left[ F^{2}|\Delta |\cos ^{2}\theta +H^{2}\sin ^{2}\theta \right] ^{1/2}.
\end{equation}
where $F$ and $G$ are given by \eqref{F} and \eqref{G}. From the denominator
of the above equation one can see that it vanishes at the horizons of the
black hole and ring singularity. Thus we study the behavior of $\Omega_{p}$ for
different values of spacetime parameters $a$ and $\alpha$ and observer's
angular velocity parameter $k$ in detail.

In FIG: (\ref{opfig}), we have plotted $\Omega_{p}$ for black hole with $a=0.5$ and $%
\alpha=1$ in left column and for naked singular with $a=1.5$ and $\alpha=1$
in right column for different observers of angular velocity parameter $%
k=0.1,0.5,0.9$ in first, second and third row, respectively. It is seen that
for a black hole the precession frequency $\Omega_{p}$ of a gyroscope
attached to any observer except ZAMO, diverges when ever it approach the
horizons along any direction. However, its remain finite for ZAMO observer
at horizon. On the other hand, for a naked singularity the precession
frequency remain finite even as the observer approach $r=0$ along any
direction except $\theta=\pi/2$. Further, if the observer approach $r=0$
along $\theta=\pi/2$ the precession frequency $\Omega_p$ becomes very large
because of the ring singularity. In FIG: (\ref{opf}), we further illustrate that for
all other choices of the parameters, the behavior of the precession
frequency $\Omega_{p}$ remain the same as in case of the black hole with $%
a=0.5$ and $\alpha=1$ and $a=1.5$ and $\alpha=1$ in naked singularity. That
is, for black hole with any $a$ and $\alpha$ the precession frequency $%
\Omega_{p}$ of gyroscope of all the observer except ZAMO, diverges near the
horizons while for naked singularity it remain finite up to $r=0$ along any
direction except $\theta=\pi/2$.

Finally, using the spin precession frequency of a test gyroscope attached to
stationary observer, we can differentiate a black hole from naked
singularity. The four velocity of an observer in the spacetime of the line
element \eqref{LE} is timelike if azimuthal components of the velocity
(equal to angular velocity) $\Omega$ at fixed $(r,\theta)$ remain in between 
$\Omega_{-}$ and $\Omega_{+}$. Further, the angular velocity can be
parameterized $q$ such that $0<q<1$. Consider there are two observer with
different angular velocity $\Omega_{1}$ and $\Omega_{2}$ approaching the
astrophysical object in the PFDM of line element %
\eqref{LE} along the different directions $\theta_1$ and $\theta_2$. If (a)
the precession frequency $\Omega_p$ of a test gyroscope of at least one
observer becomes arbitrary large as the observer approach the central object
along any direction then the object is a black hole and (b) if the
precession frequency of any of the observer along at most one of the two
directions becomes arbitrarily larges as it approach the central object,
then the object is a naked singularity.\\

\section{Observational aspects}

Experimental observations obtained with the Rossi X-ray Timing Explorer (RXTE) reveals the phenomena of quasi-periodic oscillations (QPOs) by analyzing the power spectrum of the time series 
of the X-rays  \cite{stella1,stella2,xray2,Revnew}.
%Studies show that the observed X-ray flux originates  from the accreting stellar mass near black holes and neutron stars.
There are various sources of cosmic X-rays and one of them is accreting stellar mass near compact objects like black holes and neutron stars.
A careful monitoring identifies two types of QPOs, namely the high-frequency quasi-periodic oscillations (HF QPOs), and the low-frequency
(LF QPOs). Although the theoretical explanation behind this effect is not
yet well understood, QPOs are often linked with the relativistic precession
of the accretion disk near black holes or neutron stars. QPOs may be potentially a useful tool in astrophysics for investigating new  features related to the accretion process near black holes. For example, within a certain model of X-ray timing measurements of QPOs can be used to estimate the spin angular momentum and the mass of the black hole which is of significant importance in astrophysics \cite{motta}.  Experimental data shows that, the observed HF QPOs belong to the interval $50 - 450$ Hz. Furthermore, there are three classes of LF QPOs known as: type-A, type-B, and type-C, LF QPOs, respectively. The
typical frequencies belong to the interval: $6.5-8 $Hz, $0.8-6.4$ Hz, and $0.01-30$ Hz,
respectively.
\begin{figure}[!ht]
	\centering
	%\captionsetup{justification=centering} 
	\minipage{0.50\textwidth} %
	\includegraphics[width=8.2cm,height=5.6cm]{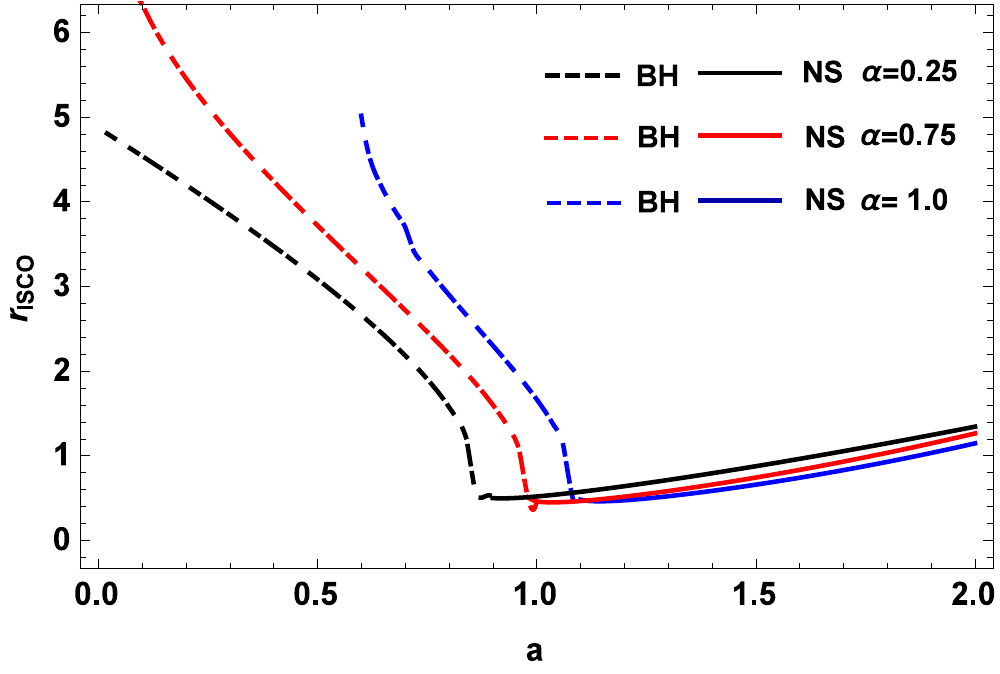}\newline 
	(a) \label{ISCOpstv} \endminipage\hfill \minipage{0.50\textwidth} %
	\includegraphics[width=8.2cm,height=5.6cm]{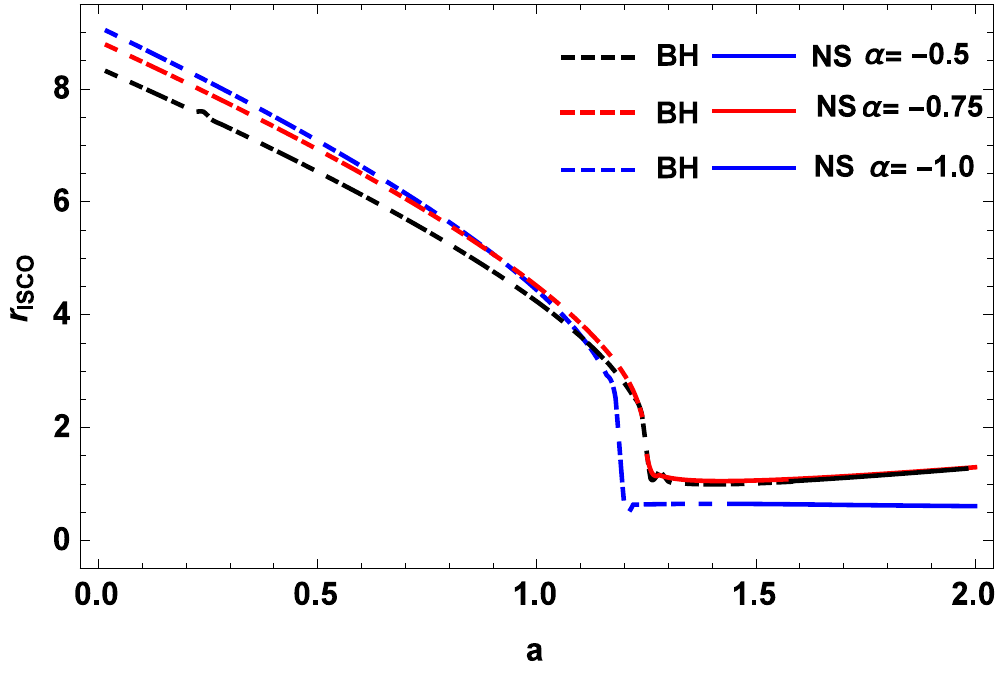}\newline
	(b) \label{ISCOngtv} \endminipage\hfill
	\caption{{\protect\footnotesize The ISCO of the black hole (BH) and naked singularity (NS) (in the unit of $M$) for positive and negative PFDM parameter $\alpha$ is plotted in panel (a) and (b), respectively. For $\alpha=-0.5,-0.75,-1.0$, the critical values of spin parameter are $a_c=1.25486557,1.246502349,1.190826462$ and for positive values of $\alpha$ the critical value of $a_c$ is obtained from \eqref{ac}.}}
	\label{ISCO}
\end{figure}
\begin{figure}[!ht]
	\centering
	%\captionsetup{justification=centering}
	 \minipage{0.50\textwidth} %
	\includegraphics[width=8.2cm,height=5.6cm]{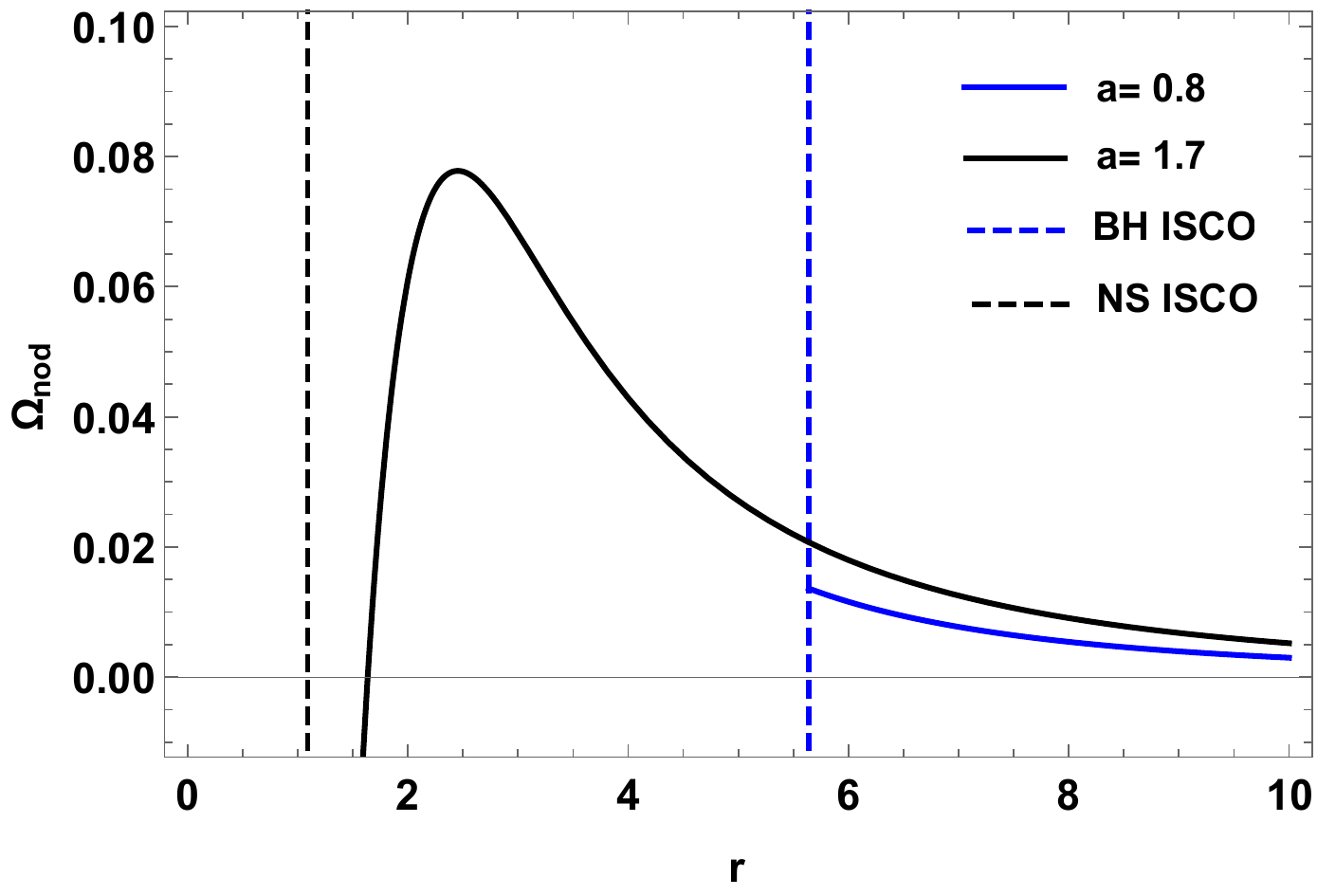}\newline
	(a)  \label{Onodngtv} \endminipage\hfill \minipage{0.50\textwidth} %
	\includegraphics[width=8.0cm,height=5.6cm]{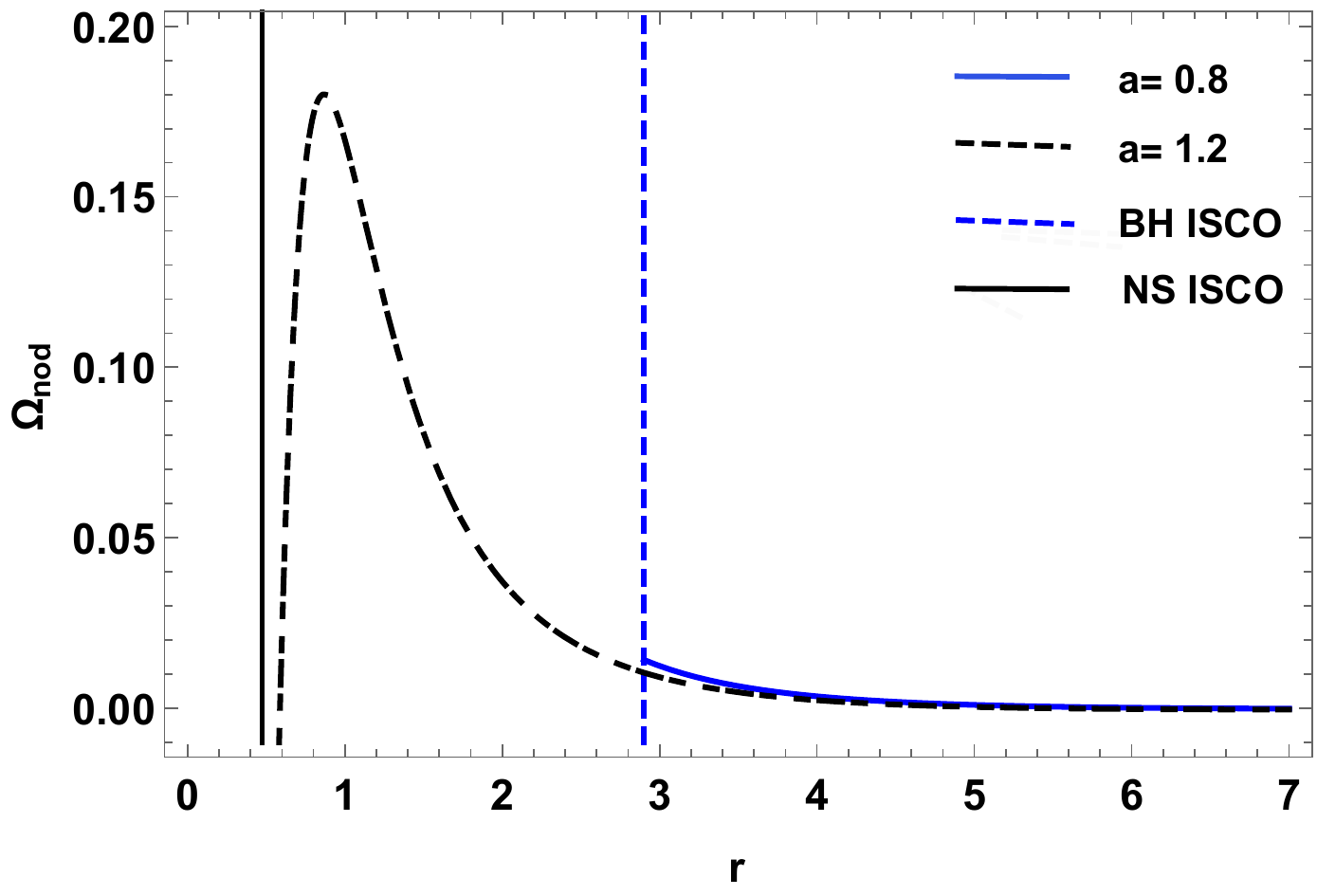}\newline
	(b) \label{Omganodpstv} \endminipage\hfill
	\caption{{\protect\footnotesize  We plot $\Omega_{nod}$ (in units of $M^{-1}$) as a function of $r$ (in units of $M$) with $\alpha=-1$ and $\alpha=1$. As can be seen, for black holes the NPPF $\Omega_{nod}$
decreases as we increase $r$. From the plots we can further observe that a naked singularity is characterized by a peak value of $\Omega_{nod}$. Furthermore $\Omega_{nod}$ vanishes at some radius $r_0$. The negative values of $\Omega_{nod}$, physically can be interpreted as a reversion of the precession direction.}}
	\minipage{0.50\textwidth} %
	\includegraphics[width=8.2cm,height=5.6cm]{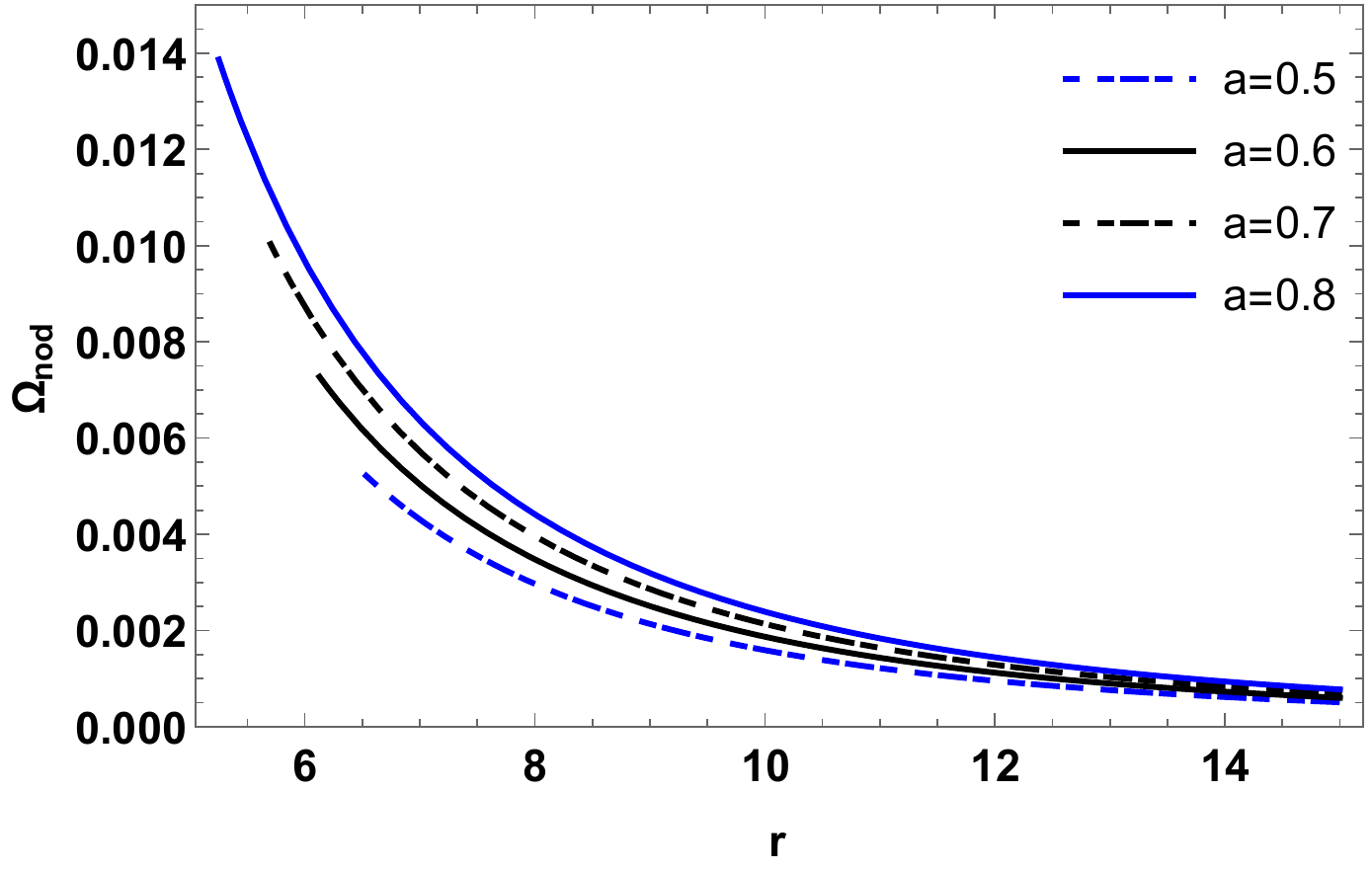}\newline
	(c)  \label{Onodngtv5} \endminipage\hfill \minipage{0.50\textwidth} %
	\includegraphics[width=8.0cm,height=5.6cm]{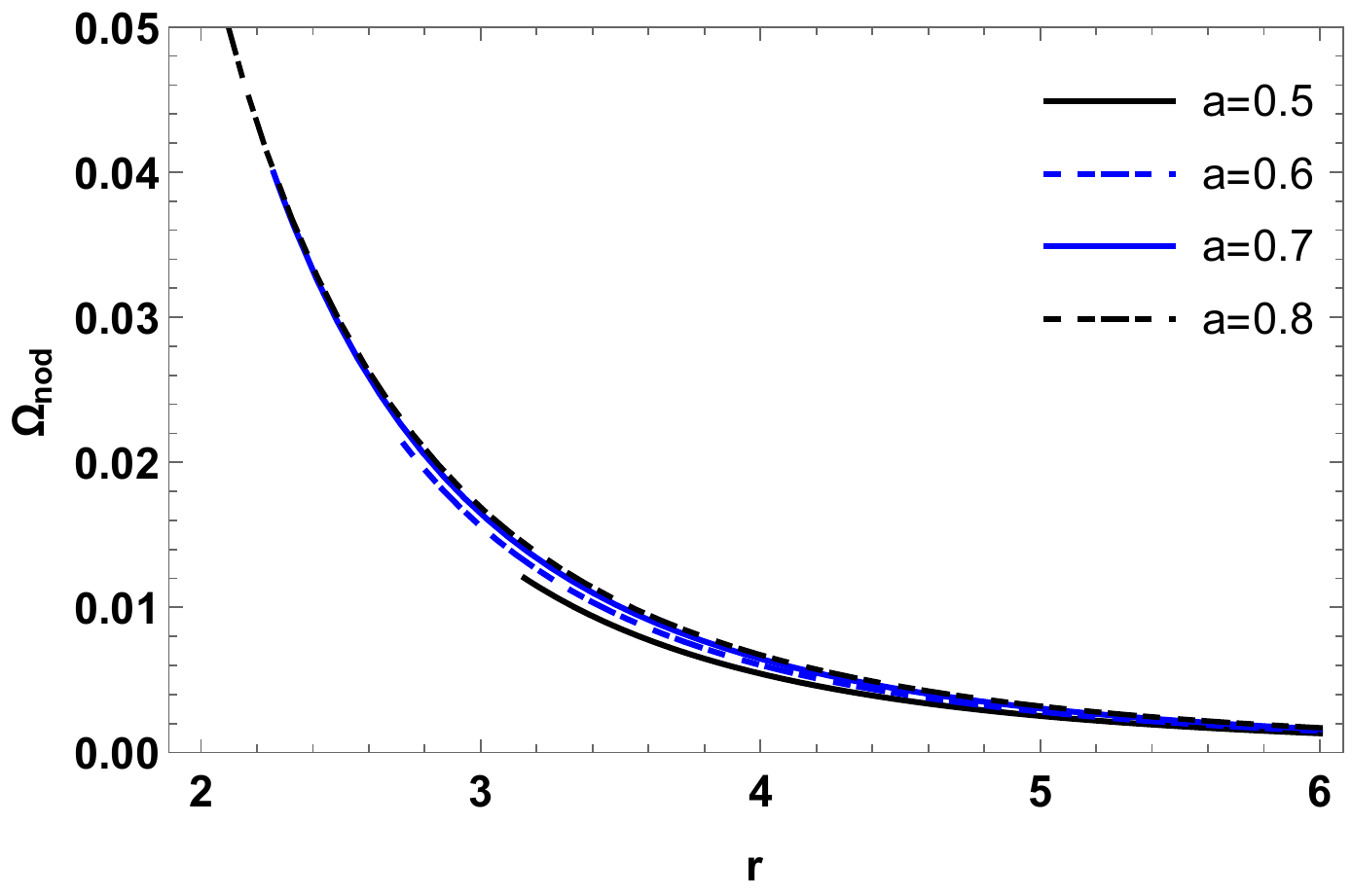}\newline
	(d) \label{Omganodpstv5} \endminipage\hfill
	\caption{{\protect\footnotesize We plot the NPPF $\Omega_{nod}$ as a function of $r$ for the black hole case with $\alpha=-0.5$ and $\alpha=0.5$. Clearly $\Omega_{nod}$ always decreases with the increase of $r$. }}
		\minipage{0.50\textwidth} %
	\includegraphics[width=8.2cm,height=5.6cm]{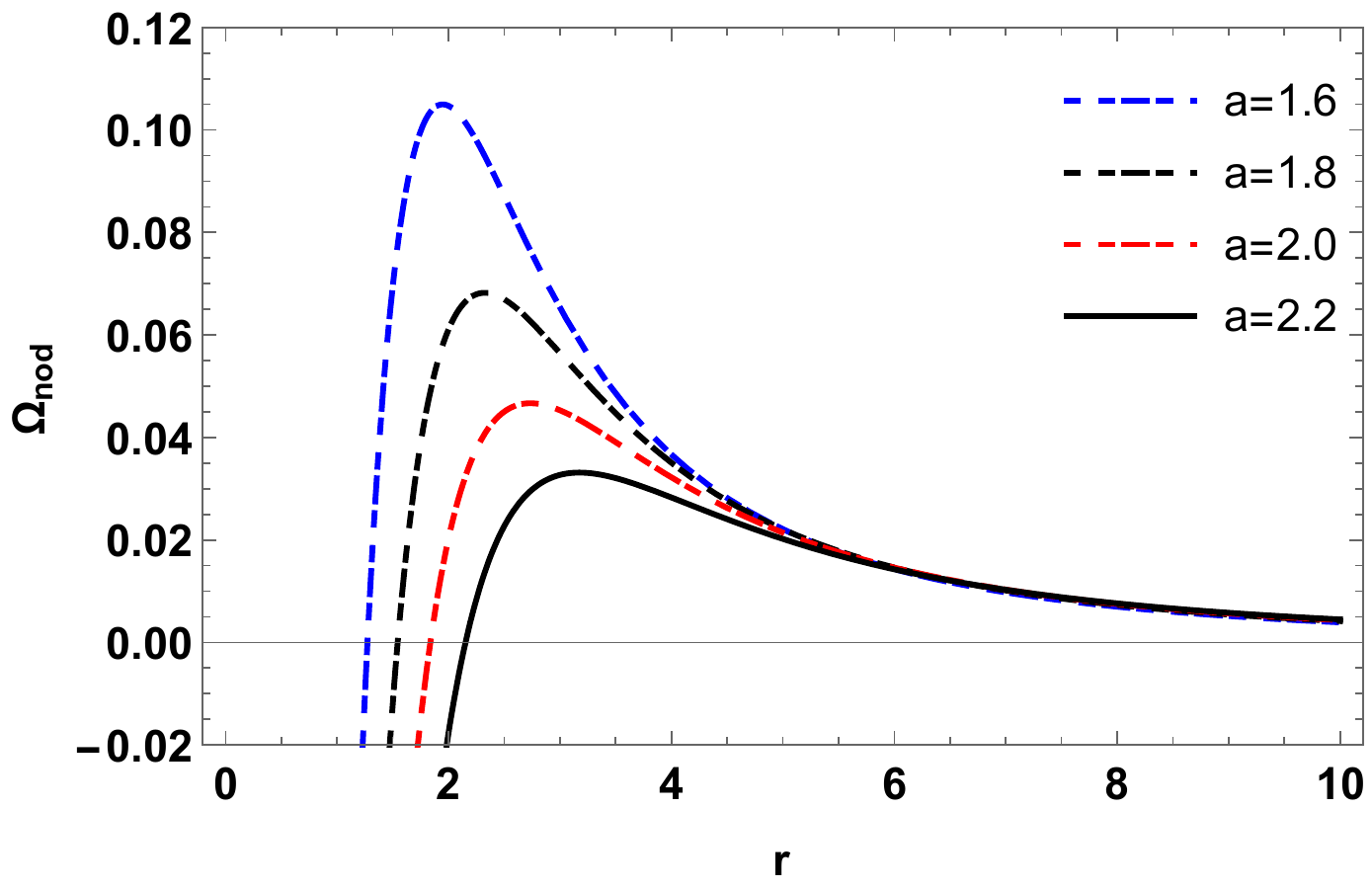}\newline
	(e) \label{Onodngtv51} \endminipage\hfill \minipage{0.50\textwidth} %
	\includegraphics[width=8.0cm,height=5.6cm]{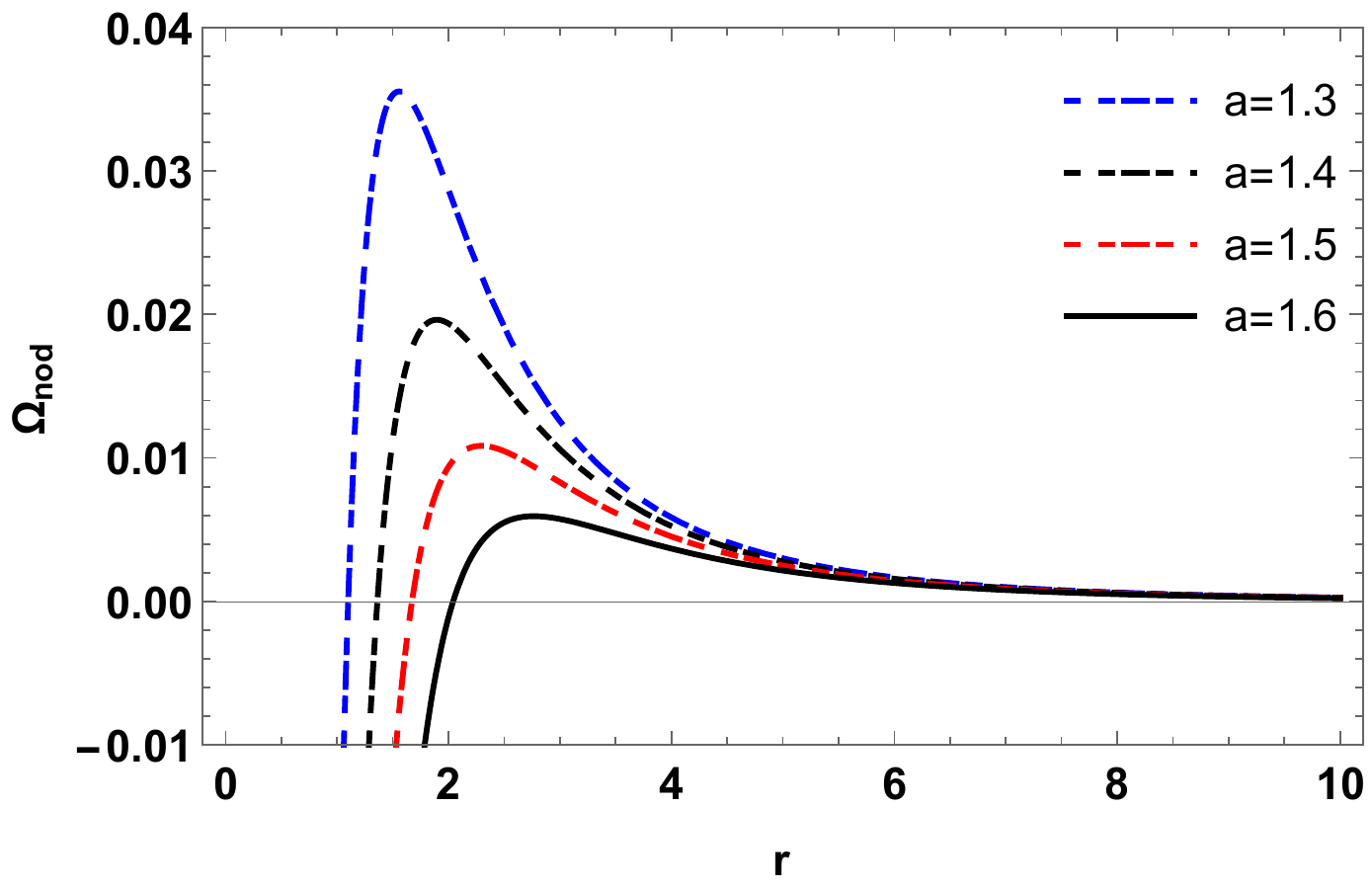}\newline
	(f) \label{Omganodpstv52} \endminipage\hfill
	
	\caption{{\protect\footnotesize The $\Omega_{nod}$ as a function of $r$ is plotted in the case of naked singularities with $\alpha=-0.5$ and $\alpha=0.5$. From the plots we can see that $\Omega_{nod}$ increases initially, then a particular peak value is obtained, and finally decreases with the increase of $r$.  The negative values of $\Omega_{nod}$, shows that the precession direction has changed. }}
	
\end{figure}\label{fig8}
\begin{figure}[!ht]
	\centering
	%\captionsetup{justification=centering}
	 \minipage{0.50\textwidth} %
	\includegraphics[width=8.2cm,height=5.6cm]{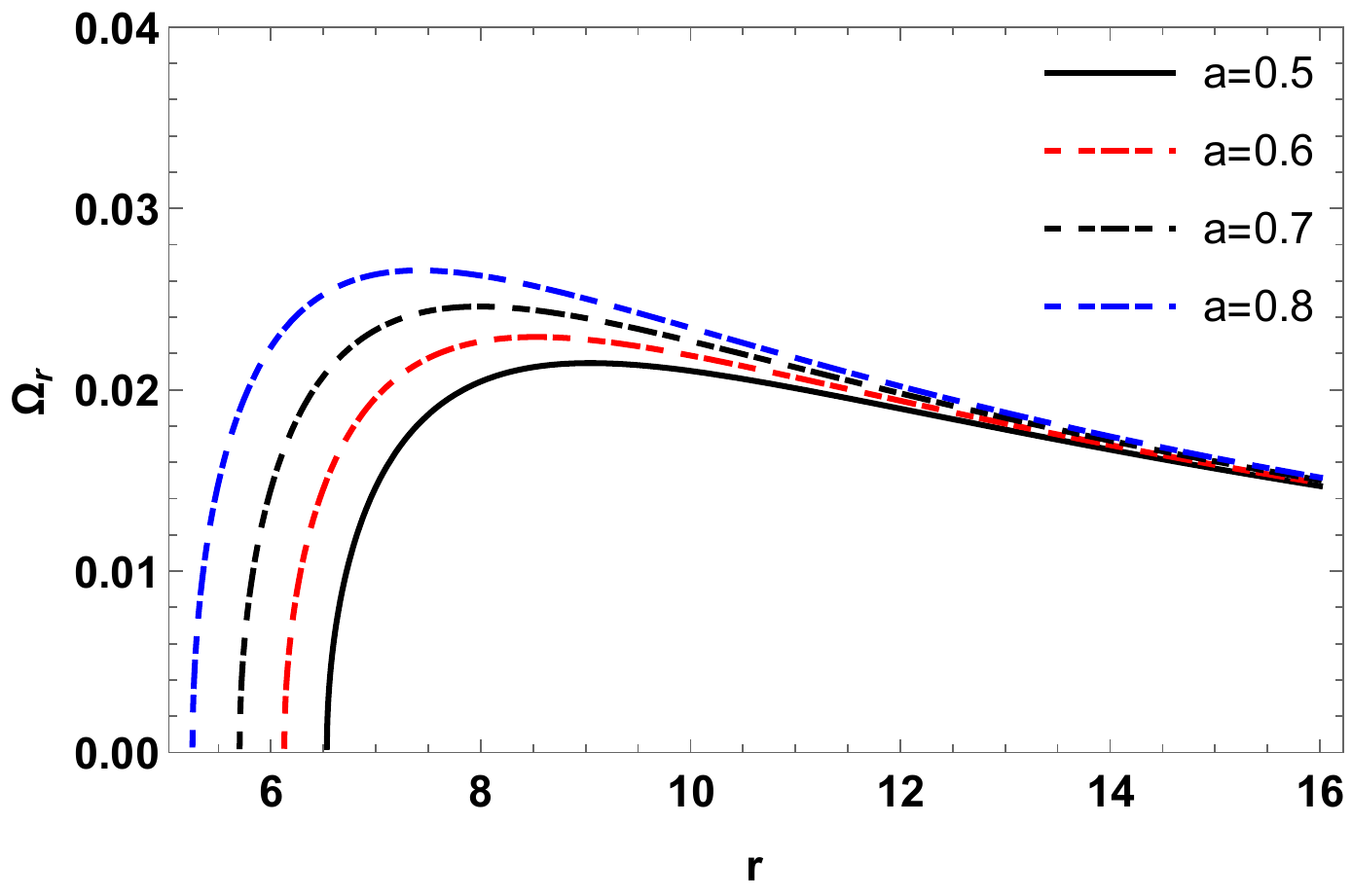}\newline
	(a)  \label{Onodngtvv} \endminipage\hfill \minipage{0.50\textwidth} %
	\includegraphics[width=8.0cm,height=5.6cm]{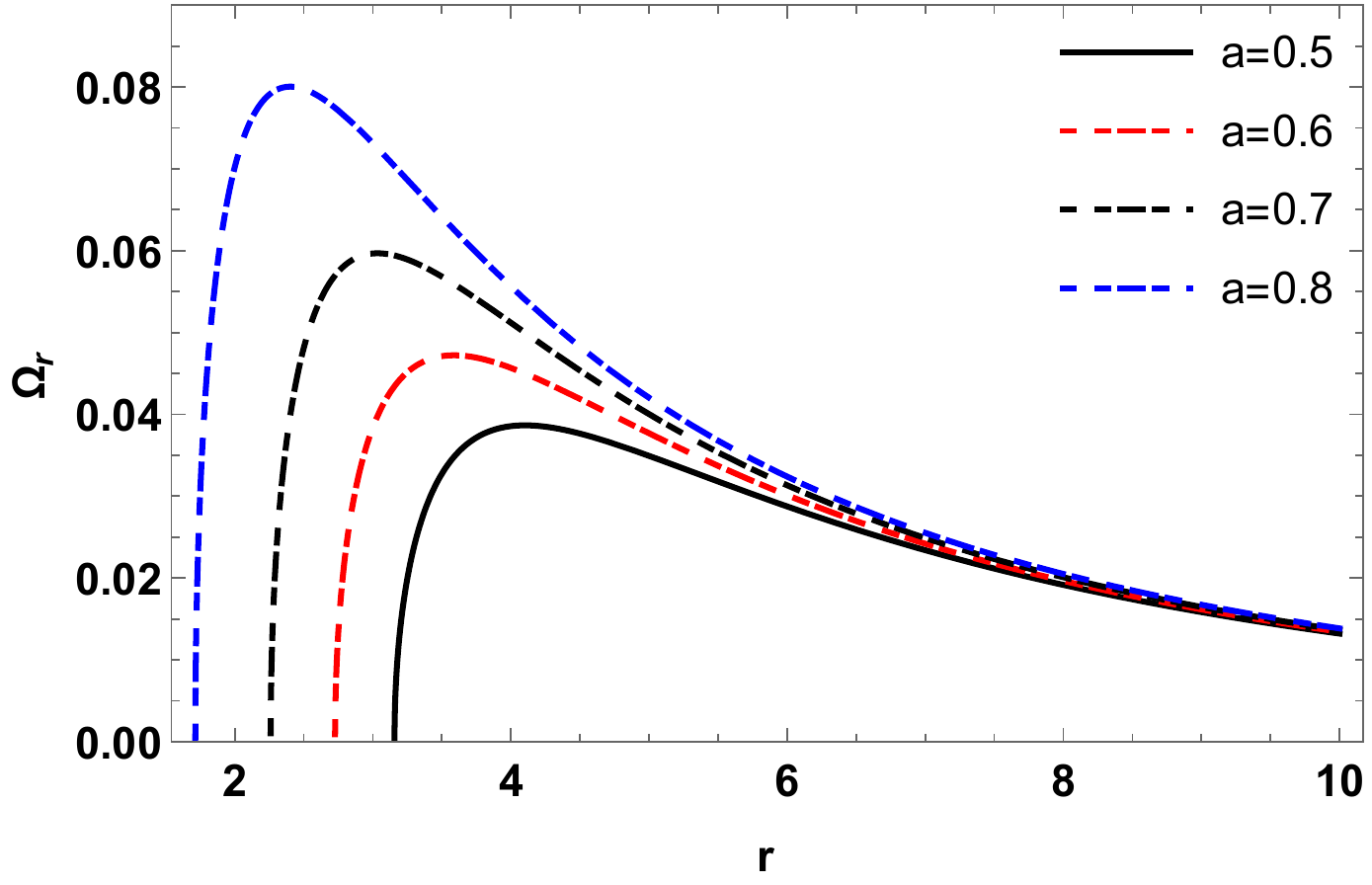}\newline
	(b) \label{Omgar} \endminipage\hfill
	\minipage{0.50\textwidth} %
	\includegraphics[width=8.2cm,height=5.6cm]{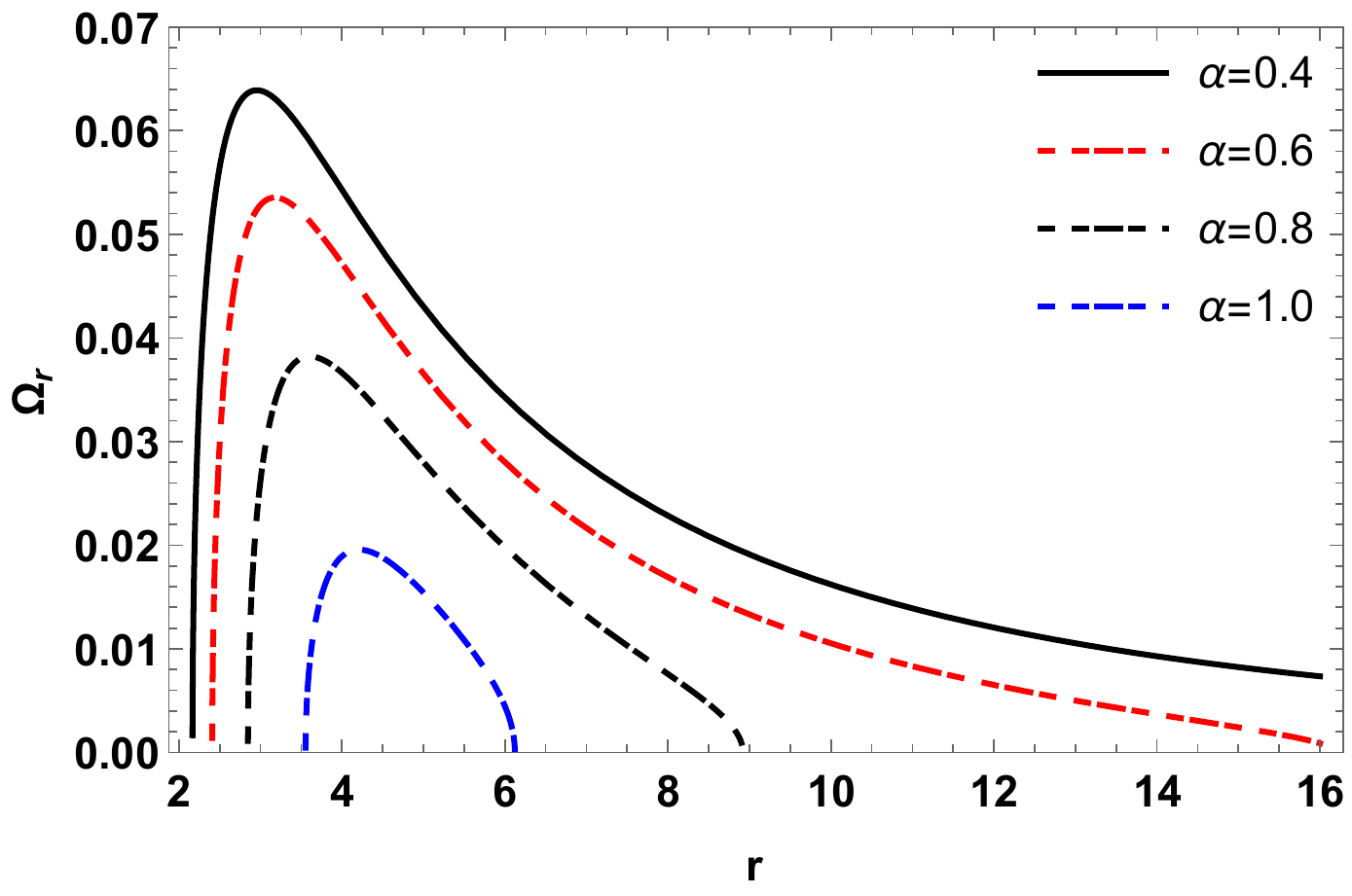}\newline
	(c) \label{Onodr2} \endminipage\hfill \minipage{0.50\textwidth} %
	\includegraphics[width=8.0cm,height=5.6cm]{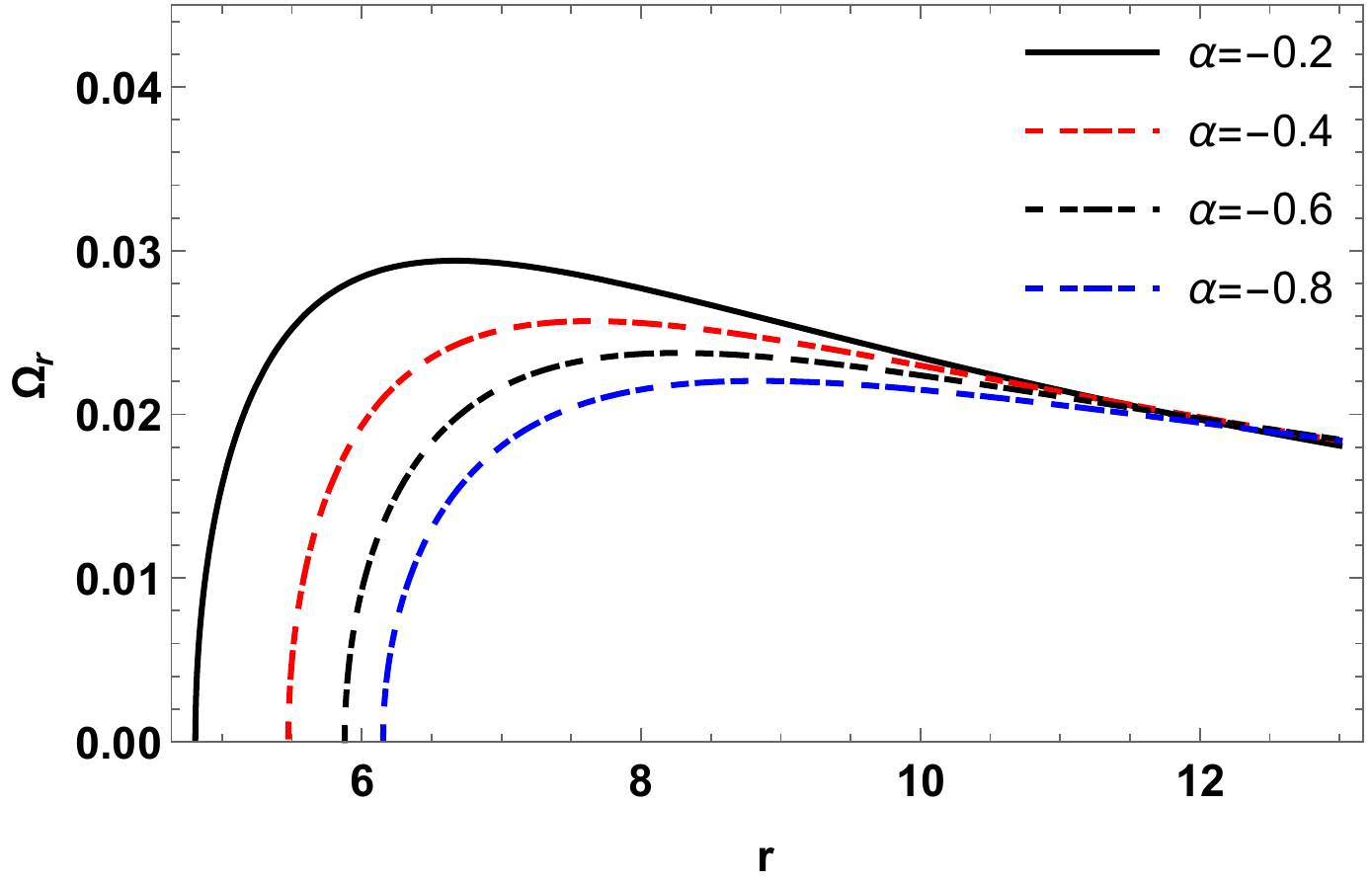}\newline
	(d) \label{Omgar3} \endminipage\hfill
	
	\caption{{\protect\footnotesize We plot $\Omega_{r}$ versus $r$ with PFDM parameter values (upper panel) $\alpha=-1$ and $\alpha=1$, and spin parameter (lower panel)  $a=-0.7$ and $a=0.7$. We can see that $\Omega_{r}$ vanishes at some particular value of $r$ depending on the particular value of PFDM parameter. Note that, $\Omega_{r}$ reaches a particular peak value, then decreases with the increase of $r$.  }}

\end{figure}\label{fig0}

\begin{table}[tbp]
\begin{tabular}{|l|l|l|l|l|l|l|l|l|l|l|l|}
\hline
 \multicolumn{2}{|c|}{  } &  \multicolumn{3}{c|}{  $\alpha=1$ (in $M$) } & \multicolumn{3}{c|}{  $\alpha=0.5$ (in $M$) } & \multicolumn{3}{c|}{  $\alpha=0$ (in $M$) }\\\hline
   $a/M$ & $r$ (in $M$) & $\nu^{\alpha}_{\phi}$ (in Hz)& $\nu^{\alpha}_{\theta}$
(in Hz) &  $\nu^{\alpha}_{\text{nod}}$ (in Hz)& $\nu^{\alpha}_{\phi}$ (in Hz) & $\nu^{\alpha}_{\theta}$ (in Hz)& $\nu^{\alpha}_{\text{nod}}$
(in Hz)  & $\nu^{0}_{\phi}$ (in Hz)& $\nu^{0}_{\theta}$
(in Hz) &  $\nu^{0}_{\text{nod}}$
(in Hz) \\ \hline
0.1 & 5.67 & 220 & 217 & 3 & 226 & 223 & 3   & 234 & 231 & 3  \\ \hline
0.2 & 5.45 & 232 & 226 & 6 & 238 & 231 & 7  & 246 & 239 & 7 \\ \hline
0.3 & 5.32 & 239 & 229 & 10 & 245 & 234 & 10  & 253 & 242 & 11\\ \hline
0.4 & 4.61 & 192 & 273 & 19 & 299 & 279 & 20   & 309 & 287 & 22 \\ \hline
0.5 & 4.30 & 320 & 293 & 27 & 333 & 302 & 31   & 344 & 310 & 34 \\ \hline
0.6 & 3.82 & 375 & 331 & 44 & 383 & 336 & 47   & 395 & 342 & 53 \\ \hline
0.7 & 3.45 & 527 & 363 & 64 & 435 & 366 &   69 & 448 & 371 & 77 \\ \hline
0.8& 2.85 & 544 & 428 & 116 & 553 & 428 & 125   & 567 & 429 & 138\\ \hline
0.9 & 2.32 & 693 & 491 & 202 & 702 & 483 & 219   & 718 & 472 & 246 \\ \hline
0.98 & 1.85 & 884 & 544 & 340  & 893 & 521 & 372 & 910 & 485 & 425  \\ \hline
0.99 & 1.45 & 1137 & 575 & 562 & 1146 & 520 & 626   & 1163 & 420 & 743 \\ \hline
0.9999 & 1.20 & 1350 & 593 & 757 & 1358 & 498 & 860   & 1375 & 276 & 1099\\ \hline
0.999999 & 1.05 & 1509 & 635 & 874 & 1517 & 507 & 1010 & 1533 & 89 & 1444 \\ \hline
1.0 & 1 & 1568 & 669 & 899 & 1575 & 532 & 1043   & 1591 & 0 & 1591\\ \hline
1.001 & 0.95 & 1629 & 724 & 905 & 1636 & 583 & 1053 & 1652 & 133 & 1519  \\ \hline
1.01 & 0.80 & 1825 & 1093 & 732 & 1830 & 968 &   862 & 1844 & 679 & 1165 \\ \hline
1.02 & 0.75 & 1888 & 1339 & 549 & 1954 & 1428 & 526 & 1967 & 1199 & 768 \\ \hline
1.04 & 0.65 & 2020 & 2025 & -5 & 2024 & 1941 & 83   & 2035 & 1753 & 282  \\ \hline
1.08 & 0.667 & 1944 & 2127 & -183 & 1948 & 2054 &   -106 & 1959 & 1894 & 65 \\ \hline
1.2 & 0.8 & 1646 & 1863 & -217 & 1650 & 1808 & -158 & 1662 & 1696 & -34  \\ \hline
2 & 1.26 & 919 & 1619 & -700 &  923 & 1609 & -686 & 932 &1588 & -656 \\ \hline
3 & 3.20 & 353 & 453 & -100 & 357 & 453 & -96   & 365 & 453 & -88 \\ \hline
4 & 4.00 & 256 & 371 & -115 &260 & 373 &-113 & 265 & 375 & -110 \\ \hline
\end{tabular}
\begin{tabular}{|l|l|l|l|l|l|l|l|l|l|l|l|}
\hline
 \multicolumn{2}{|c|}{  } &  \multicolumn{3}{c|}{  $\alpha=-1$ (in $M$) } & \multicolumn{3}{c|}{  $\alpha=-0.5$ (in $M$) } & \multicolumn{3}{c|}{  $\alpha=-0.1$ (in $M$) }\\\hline
   $a/M$ & $r$ (in $M$) & $\nu^{\alpha}_{\phi}$ (in Hz)& $\nu^{\alpha}_{\theta}$
(in Hz) &  $\nu^{\alpha}_{\text{nod}}$ (in Hz) & $\nu^{\alpha}_{\phi}$ (in Hz) & $\nu^{\alpha}_{\theta}$ (in Hz)& $\nu^{\alpha}_{\text{nod}}$
(in Hz)  & $\nu^{\alpha}_{\phi}$ (in Hz)& $\nu^{\alpha}_{\theta}$
(in Hz) &  $\nu^{\alpha}_{\text{nod}}$
(in Hz) \\ \hline
0.1 & 5.67 & 247 & 243 & 4 & 242 & 238 & 4   & 236 & 233 & 3  \\ \hline
0.2 & 5.45 & 259 & 251 & 8 & 254 & 246 & 8  & 248 & 241 & 7 \\ \hline
0.3 & 5.32 & 267 & 254 & 13 & 261 & 249 & 12  & 255 & 244 & 11\\ \hline
0.4 & 4.61 & 325 & 299 & 26 & 318 & 294 & 24   & 312 & 289 & 23 \\ \hline
0.5 & 4.30 & 354 & 317 & 37 & 348 & 312 & 35   & 341 & 307 & 34 \\ \hline
0.6 & 3.82 & 412 & 352 & 60 & 406 & 349 & 57   & 398 & 344 & 54 \\ \hline
0.7 & 3.45 & 467 & 378 & 89 & 460 & 376 & 84 & 451 & 372 & 79 \\ \hline
0.8& 2.85 & 589 & 427 & 162 & 581 & 429 & 152   & 571 & 429 & 141\\ \hline
0.9 & 2.32 & 741 & 449 & 292 & 733 & 460 & 273   & 722 & 469 & 253 \\ \hline
0.98 & 1.85 & 935 & 410 & 525  & 926 & 443 & 483 & 925 & 474 & 441  \\ \hline
0.99 & 1.45 & 1188 & 113 & 1075 & 1180 & 281 & 899 & 1169 & 387 & 782 \\ \hline
0.9999 & 1.60 & 1070 & 315 & 762 & 1069 & 384 & 685   & 1058 & 455 & 613\\ \hline
0.999999 & 1.50 & 1147 & 224 & 923 & 1138 & 330 & 808 & 1127 & 415 & 1712 \\ \hline
1.0 & 1.55 & 1111 & 277 & 834 & 1138 & 330 & 808   & 1089 & 432 & 660\\ \hline
1.001 & 1.95 & 879 & 423 & 455 & 870 &450 & 420 & 859 & 475 & 384  \\ \hline
1.01 & 1.80 & 954 & 398 & 555 & 945 & 436 &   509 & 934 & 472 & 462 \\ \hline
1.02 & 1.75 & 978 & 386 & 592 & 970 & 429 & 541 & 959 & 470 & 489 \\ \hline
1.04 & 1.65 & 1032 & 357 & 675 & 1023 & 413 & 610   & 1012 & 464 & 548  \\ \hline
1.08 & 1.667 & 1008 & 378 & 630 & 1000 & 430 & 570 & 980 & 478 & 512 \\ \hline
1.2 & 1.8 & 902 & 437 & 465 & 895 & 472 & 422 & 985 & 506 & 378  \\ \hline
2 & 1.26 & 944 & 1550 & -606 &  941 & 1565 & -624 & 935 &1518 &-646  \\ \hline
3 & 3.20 & 376 & 449 & -73 & 372& 451 & -79   & 367 & 457 & -85 \\ \hline
4 & 4.00 & 274 & 377 & -103 &271 & 377 &-106 & 267 & 375 & -108 \\ \hline
\end{tabular}
\caption{We have considered an object with mass $M=10 {M}_\odot$ to
calculate the KF precession   $%
\protect\nu^{\protect\alpha}_{\theta}$, VEF precession $%
\protect\nu^{\protect\alpha}_{\theta}$,  and the NPPF  $%
\protect\nu^{\protect\alpha}_{\text{nod}}$, respectively. We have chosen the ISCO radius $r=r_{ISCO}$ given by the interval $M\leq r_{ISCO} \leq 6 M $.}
\end{table}

There are three characteristic frequencies attributed to a test particle orbiting the black hole, namely the KF $\Omega _{\phi } $, the VEF $%
\Omega _{\theta }$, and the radial epicyclic frequency (REF)  $%
\Omega _{r}$, respectively. By taking into account the effect of PFDM we have calculated the following expressions for the characteristic frequencies \cite{kepfreq}:
\begin{eqnarray}
\Omega _{\phi } &=&\pm \frac{\sqrt{M+\frac{\alpha }{2}\left( 1-\ln \left( 
\frac{r}{|\alpha |}\right) \right) }}{r^{3/2}\pm a\sqrt{M+\frac{\alpha }{2}%
\left( 1-\ln \left( \frac{r}{|\alpha |}\right) \right) }}, \\
\Omega _{r} &=&\frac{\Omega _{\phi }}{r}\left[ r^{2}-6Mr-3a^{2}-\alpha
r\left\{ 1-3\ln \left( \frac{r}{|\alpha |}\right) \right\} \pm 8a\sqrt{rM+%
\frac{\alpha r}{2}\left\{ 1-\ln \left( \frac{r}{|\alpha |}\right) \right\} }-%
\frac{\alpha \left\{ a^{2}+ar+r^{2}\right\} }{\left\{ 2M+2\alpha \left(
1-\ln \left( \frac{r}{|\alpha |}\right) \right) \right\} }\right] ^{\frac{1}{%
2}}, \\
\Omega _{\theta } &=&\frac{\Omega _{\phi }}{r}\left[ r^{2}+3a^{2}\mp 4a\sqrt{%
rM+\frac{\alpha r}{2}\left\{ 1-\ln \left( \frac{r}{|\alpha |}\right)
\right\} }-\frac{a\alpha \left\{ a\mp 2\sqrt{rM+\frac{\alpha r}{2}\left\{
1-\ln \left( \frac{r}{|\alpha |}\right) \right\} }\right\} }{\left\{ M+\frac{%
\alpha }{2}\left( 1-\ln \left( \frac{r}{|\alpha |}\right) \right) \right\} }%
\right] ^{\frac{1}{2}}.
\end{eqnarray}

One can easily show that in the limiting case when $\alpha$ vanishes, the characteristic frequencies in the Kerr geometry are obtained \cite{NS,mottanew,bambi}. With these results in mind, we can extract further informations by defining the following two quantities 
\begin{equation}
\Omega_{\text{nod}}=\Omega _{\phi }-\Omega _{\theta },
\end{equation}
and 
\begin{equation}
\Omega_{\text{per}}=\Omega _{\phi }-\Omega _{r }.
\end{equation}
Where $\Omega_{\text{nod}}$ measures the orbital plane precession and is usually known as the NPPF (or Lense-Thirring precession frequency), on the other hand $\Omega_{\text{per}}$ measures the precession of the orbit and is known as the periastron precession frequency. From FIG. (9)-(11) we can see that in the black hole case, the NPPF $\Omega_{nod}$ always decreases with $r$, therefore we can write the following condition
\begin{equation}
\frac{d\Omega_{nod}}{dr}<0.
\end{equation}

On the other hand, an interesting feature arises in the case of naked singularities, namely $\Omega_{nod}$ initially increases, then a particular peak value is recovered, and finally $\Omega_{nod}$ decreases with the increase of $r$. Therefore, in the case of naked singularities we can have the following condition when $\Omega_{nod}$ increases with with $r$, written as follows
\begin{equation}
\frac{d\Omega_{nod}}{dr}>0.
\end{equation}

Finally we point out that negative values of $\Omega_{nod}$, can be interpreted as a reversion of the precession direction.  

We provide a detailed analyses of our results in Table I, where we highlight the observational aspects by calculating the impact of the PFDM parameter on different frequencies: KF $%
\protect\nu^{\protect\alpha}_{\theta}$, VEF $%
\protect\nu^{\protect\alpha}_{\theta}$,  and the NPPF  $%
\protect\nu^{\protect\alpha}_{\text{nod}}$. Our results reveal that, typical values of the PFDM parameter $\alpha$ significantly affects these frequencies. In particular we find that, with the increase of positive $\alpha $, all frequencies become smaller and smaller.
On the other hand, with the decrease of negative $\alpha $, all frequencies become bigger and bigger.  
Our results further indicate that, the effects of PFDM are getting stronger with the increase of spin angular momentum parameter $a$. From the Table I, we see that one can identify LF QPOs with $\nu _{\text{nod}}^{\alpha }$, for a slowly rotating black holes, i.e $a/M<0.5$. A significant difference between $%
\protect\nu^{\protect\alpha}_{\text{nod}}$ and $%
\protect\nu^{0}_{\text{nod}}$, occurs when $a/M>0.5$. Clearly, in this range, we can identify $%
\protect\nu^{\protect\alpha}_{\text{nod}}$ with
HF QPOs.  
%In addition we emphasis that the obtained frequency range lies in the observed typical QPOs and hence are consistent with observations. 
In the near future we plan to investigate the relativistic precession model to get constraints on $\alpha$ using the data of GRO J1655-40 and by following the analysis presented by Bambi \cite{bambi}. 
%We expect that future experiments could produce better constraints for the PFDM parameter$%
%\alpha $.

\section{Conclusion}

In this paper, we have studied rotating object in PFDM %
\eqref{LE} to differentiate a Kerr-like black hole from naked singularity.
For a black hole we find the lower bound of the dark matter parameter $%
\alpha $, $-2\leq \alpha$ and gives the critical value of spin parameter $%
a_c $. For any fixed chosen $\alpha$ the line element \eqref{LE} represents
a black hole with two horizons if $a<a_c$, extremal black hole with one
horizon if $a=a_c$ and naked singularity if $a>a_c$. It is seen that for $%
\alpha< 0$, $a_c$ has maxima at $\alpha=\overline\alpha$ and for $\alpha>0$
it has minima at $\alpha=\tilde{\alpha}$. Further, for large value of $%
\alpha $, $a_c$ increases very large without any limit and thus a highly
spinning black hole can form. We also study the horizon of extremal black
hole and find that for $\alpha<0$ the size of extremal horizon $r_e$
decrease whereas for $\alpha>0$, $r_e$ has minima at $\alpha=2/3$. Further,
as $r_-\leq r_e \leq r_+$, so we can conclude that for any fixed $a$ and $%
-2\leq \alpha < 2/3 $, with increasing $\alpha$, size of the black hole
horizons decreases while for $2/3<\alpha$ increases.\\

We also studied the spin precession frequency $\Omega_p$ of the a test
gyroscope attached to a stationary timelike observer. For timelike observer
we find the restricted domain for the angular velocity $\Omega$ of the
observer. From the precession frequency $\Omega_{p}$, by setting $\Omega=0$,
we obtained LT- precession frequency for a static observer which
can lies only outside the ergosphere. We also find the geodetic precession
frequency which is due to the of spacetime curvature of Schwarzschild black
hole in PFDM. It is seen that for $\alpha<0$ the
geodetic precession frequency is increases while for $\alpha>0$ is increases.

Using spin precession frequency criteria we differentiate a black hole from
a naked singularity. We parameterize the angular velocity of the observer
and studied the spin precession frequency for different choices of the
parameter along the different directions. If the precession frequency of a
gyroscope attached to at least one of two observers with different angular
velocities blow up if they approach the central object in PFDM along any direction then the object is a black hole. If the
precession frequency of all the observer show divergence as the observers
approach the center of spacetime along at most one direction then it is
naked singularity. 

We have summarized our results by computing the effect of PFDM on the KF, VEF, and NPPF given in Table I.  We observe that frequencies depend upon the value of $a/M$, radial distance $r$, as well as the PFDM parameter $\alpha$,  yielding notable differences in the corresponding frequencies of black holes and naked singularities. We have shown that, with the increase of positive PFDM parameter, all frequencies become smaller. Consequently, with the decrease of negative PFDM parameter, all frequencies become bigger. Following our results we can conclude that LF QPOs can be identified for $a/M<0.5$. However, most significant changes are observed in the interval $a/M>0.5$, whose frequencies can be identified with HF QPOs. 

Given the fact that the accretion disk changes with time, say, when the accretion disk approaches the black hole/naked singularity, we need to study the evolution of QPO frequencies to distinguish black holes from naked singularities. In particular, for a given value of PFDM parameter $\alpha$ and $a/M$, if the accretion disc approaches the $r_{ISCO}$, we see that $\Omega_{nod}$ always increases reaching its maximum value. Interestingly, in the case of naked singularities, if the accretion disc approaches the $r_{ISCO}$, we find that $\Omega_{nod}$ firstly increases, then reaches its peak, and finally decreases. In fact, contrary to the black holes, in the case of naked singularity $\Omega_{nod}$ can be zero, as can be seen from FIG. (9). Finally, we anticipate that future experiments could produce constraints for the PFDM parameter $\alpha $.  For example, recently Bhattacharyya  has proposed a model of fast radio bursts from neutron stars plunging into black hole that implies the existence of event horizon, LT effects and the emission of gravitational waves from a black hole \cite{bhattacharyya}. In this context, it would certainly be interesting to explore the possible impact of the PFDM parameter $\alpha$ on the gravitational wave signatures.
\acknowledgments
This work is supported by National University of Modern
Languages, H-9, Islamabad, Pakistan (M. R). We would like to thank the referees for their valuable comments which helped to improve the manuscript.

\end{document}